\documentclass{aa}
\usepackage{graphicx}
\usepackage{supertabular}
\usepackage{natbib}
\usepackage{multirow}
\bibpunct{(}{)}{;}{a}{}{,}% to follow the A&A style
\usepackage{longtable,lscape}
\usepackage[varg]{txfonts}
\usepackage{pdfpages}
\usepackage{psfig}

\usepackage{color}
\begin{document}
 \title{Multiwavelength observations of GRB 140629A: \thanks{Research supported by the China Scholarship Council}\fnmsep }
 \subtitle{A long burst with an achromatic jet break in the optical and X-ray afterglow}
 \author{Y.-D. Hu\inst{1,2}
 \and S. R. Oates\inst{1,3}
 \and V. M. Lipunov\inst{4,5}
 \and B.-B. Zhang\inst{1,6,7}
 \and A. J. Castro-Tirado\inst{1,8}
 \and S. Jeong\inst{1,9}
 \and R. S\'anchez-Ram\'irez\inst{1,10}
 \and J. C. Tello\inst{1}
 \and R. Cunniffe\inst{1}
 \and E. Gorbovskoy\inst{5}
 \and M. D. Caballero-Garc\'ia\inst{1,11}
 \and S. B. Pandey\inst{12} 
 \and V. G. Kornilov\inst{4,5}
 \and N. V. Tyurina\inst{5}
 \and A. S. Kuznetsov\inst{5}
 \and P. V. Balanutsa\inst{5}
 \and O. A. Gress\inst{13}
 \and I. Gorbunov\inst{5}
 \and D. M. Vlasenko\inst{4,5}
 \and V. V. Vladimirov\inst{4,5}
 \and N. M. Budnev\inst{13}
 \and F. Balakin\inst{4}
 \and O. Ershova\inst{13}
 \and V. V. Krushinski\inst{14} 
 \and A. V. Gabovich\inst{15}
 \and V. V. Yurkov\inst{15} 
 \and J. Gorosabel\inst{1\dagger}
 \and A. S. Moskvitin\inst{16}
 \and R. A. Burenin\inst{17,18}
 \and V. V. Sokolov\inst{16}
 \and I. Delgado\inst{10,19}
 \and S. Guziy\inst{20}
 \and E. J. Fernandez-Garc\'ia\inst{1}
 \and I. H. Park\inst{9}
 }

\institute{Instituto de Astrof\'isica de Andaluc\'ia (IAA-CSIC), Glorieta de la Astronom\'ia s/n, E-18008, Granada, Spain,\\ e-mail: huyoudong072@hotmail.com.
\and Universidad de Granada, Facultad de Ciencias Campus Fuentenueva S/N CP 18071 Granada, Spain.
\and Department of Physics, University of Warwick, Coventry, CV4 7AL, UK.
\and M.V.Lomonosov Moscow State University, Physics Department, Leninskie gory, GSP-1, Moscow, 119991, Russia.
\and M.V.Lomonosov Moscow State University, Sternberg Astronomical Institute, Universitetsky pr., 13, Moscow, 119234, Russia.
\and School of Astronomy and Space Science, Nanjing University, 210093, Nanjing, China, e-mail: bbzhang@nju.edu.cn.
\and Key Laboratory of Modern Astronomy and Astrophysics, Nanjing University, Ministry of Education, Nanjing, 210093, China.
\and Unidad Asociada CSIC Departamento de Ingenier\'ia de Sistemas y Autom\'atica, Escuela de Ingenieros Industriales, Universidad de M\'alaga, C. Dr. Ortiz Ramos sn, 29071 M\'alaga, Spain.
\and Institute of Science and Technology in Space, SungKyunKwan University, Suwon 16419, Republic of Korea.
\and INAF, Istituto Astrofisica e Planetologia Spaziali, Via Fosso del Cavaliere 100, I-00133 Roma, Italy.
\and Astronomical Institute, Academy of Sciences of the Czech Republic, Bo\v{c}n\'{\i}~II 1401, CZ-141\,00~Prague, Czech Republic.
\and Aryabhatta Research Institute of Observational Sciences, Manora Peak, Nainital - 263 002, India.
\and Applied Physics Institute, Irkutsk State University, 20, Gagarin blvd, 664003, Irkutsk, Russia.
\and dourovka Astronomical Observatory, Ural Federal University, Lenin ave. 51, Ekaterinburg 620000, Russia.
\and Blagoveshchensk State Pedagogical University, Lenin str., 104, Amur Region, Blagoveschensk 675000, Russia.
\and Special Astrophysical Observatory, Russian Academy of Sciences, Nizhnii Arkhyz, 369167 Russia.
\and Space Research Institute RAS (IKI),84/32 Profsoyuznaya, Moscow 117997, Russia.
\and National Research University Higher School of Economics, Moscow, 101000, Myasnitskaya Street 20, Russia. 
\and Universidad Internacional de Valencia (VIU), Calle Pintor Sorolla 21, E-46002, Valencia, Spain.
\and Mykolaiv National Unuversity, Nikolska 24, Mykolaiv, 54030, Ukraine.
}
\date{Received 2018/ 2018}
 \abstract{}
 {We investigate the long gamma-ray burst (GRB)\,140629A through multiwavelength observations to derive the properties of the dominant jet and its host galaxy.}
 {The afterglow and host galaxy observations were taken in the optical ({\it Swift}/UVOT and various facilities worldwide), infrared ({\it Spitzer}), and X-rays ({\em Swift}/XRT) between 40 seconds and 3 yr after the burst trigger.}
 {Polarisation observations by the MASTER telescope indicate that this burst is weakly polarised. The optical spectrum contains absorption features, from which we confirm the redshift of the GRB as originating at z=2.276$\pm$0.001. We performed spectral fitting of the X-rays to optical afterglow data and find there is no strong spectral evolution. We determine the hydrogen column density $N_{H}$ to be $7.2\times 10^{21}$cm$^{-2}$ along the line of sight. The afterglow in this burst can be explained by a blast wave jet with a long-lasting central engine expanding into a uniform medium in the slow cooling regime. At the end of energy injection, a normal decay phase is observed in both the optical and X-ray bands. An achromatic jet break is also found in the afterglow light curves $\sim$0.4 d after trigger. We fit the multiwavelength data simultaneously with a model based on a numerical simulation and find that the observations can be explained by a narrow uniform jet in a dense environment with an opening angle of $6.7^{\circ}$ viewed $3.8^{\circ}$ off-axis, which released a total energy of $1.4\times 10^{54}$erg. Using the redshift and opening angle, we find GRB 140629A follows both the Ghirlanda and Amati relations. From the peak time of the light curve, identified as the onset of the forward shock (181s after trigger), the initial Lorentz factor ($\Gamma_{0}$) is constrained in the range 82-118. Fitting the host galaxy photometry, we find the host to be a low mass, star-forming galaxy with a star formation rate of log(SFR)=$1.1_{-0.4}^{+0.9}$ M$_\odot$yr$^{-1}$. We obtain a value of the neutral hydrogen density by fitting the optical spectrum, log$N_{H I}=21.0\pm0.3$, classifying this host as a damped Lyman-alpha. High ionisation lines (\ion{N}{V},\ion{Si}{IV}) are also detected in the spectrum.
 }{}
\maketitle
%%%%%%%%%%%%%%%%%%%%%%%%%%%%%%%%%%%%%%%%%%%%%%
%-----------------------------------------------------------------------------------------------------------
\section{Introduction}
Gamma-ray bursts (GRBs) are the most violent explosions in the Universe, releasing $10^{48}-10^{54}$ ergs (if considered isotropic)
typically within a few seconds in gamma rays; but these explosions have been observed up to a few hours in some instances~\citep{2014ApJ...787...66Z,2015Natur.523..189G}. Gamma-ray bursts can be divided into two classes depending on their duration: long (>2 s) and short ($\leq$2 s)~\citep{1993ApJ...413L.101K}, the progenitors of which are thought to be the collapse of massive stars or the merger of two compact objects~\citep{2004IJMPA..19.2385Z,2015PhR...561....1K}, respectively. In the final stages of merger or collapse, a highly collimated ejecta is released, which has a typical opening angle $\theta_{jet}=5^{\circ}$-$10^{\circ}$ \citep{2009ApJ...698...43R,2015ApJ...806...15Z}. An internal dissipation process within the jet is thought to produce prompt gamma-ray  emission~\citep{1994ApJ...430L..93R,1997ApJ...490...92K,1998MNRAS.296..275D,2009ApJ...707.1623M,2014ApJ...789..145H}, while a longer lived, multiwavelength afterglow is expected to be produced as the jet ploughs into the circumstellar medium \citep[of constant density or a stellar-wind-like density;][]{1997ApJ...476..232M,1997MNRAS.287..110S}. The relativistic effect implies that emission from the jet is beamed into a cone of half-opening angle 1/$\varGamma_{0}$~\citep{1997ApJ...487L...1R,2004RvMP...76.1143P,2007RMxAC..27..140G,2008ApJ...675..528L,2013ApJ...767..141V}, where $\varGamma_{0}$ is the initial Lorentz factor of the jet, typically of a value of several hundred \citep{1999PhR...314..575P}. The beamed geometry leaves a clear signature on the afterglow light curve, manifesting itself as an achromatic break known as a jet break, occurring simultaneously at all frequencies, days to weeks after the burst~\citep{1999ApJ...519L..17S,1999ApJ...525..737R}. 
This jet break, resulting in a steeper decay index, occurs when $\varGamma_{0}$ has decreased to 1/$\theta_{jet}$. The shallower decay index, observed prior to the jet break, is maintained owing to the observer receiving emission from an increasing proportion of the jet as $\varGamma$ decreases~\citep{2006ApJ...642..354Z}. Once the observer sees the entire jet, the jet break is observed. The geometry and angular size of the jet, directly affects measurements of the GRB energy and event rate. The isotropic energy should therefore be corrected by the collimation correction factor, $f_{b}$=(1-$\cos\theta_{jet}$), which solves the energy budget problem~\citep{2003ApJ...594..674B,2001ApJ...562L..55F,2005ApJ...627....1F,2008ApJ...680..531K,2009ApJ...698...43R}. Hence, the detection of a jet break in the afterglow light curve is an important diagnosis for constraining the outflow geometry and burst energetics. Although the determination of the jet opening angle from the observed break in the afterglow light curves depends on the model \citep[e.g. assumed jet structure, radiation efficiency, and circumburst matter density profile;][]{1999ApJ...519L..17S,2001ApJ...562L..55F}.

Much of our current understanding of GRB jets has been built upon observational data. Generations of facilities, including the {\it Compton Gamma-ray Observatory (CGRO)}, {\it Beppo-SAX}, {\it High Energy Transient Explorer 2 (HETE-2)}, {\it Konus}-WIND, {\it INTErnational Gamma-Ray Astrophysics Laboratory (INTEGRAL)}~\citep{1992Natur.355..143M,1995SSRv...71..265A,1997Natur.387..783C,2003AIPC..662....3R,2005A&A...438.1175R} have been used to study such catastrophic events since they were first detected almost half a century ago. In particular, the {\it Neil Gehrels Swift Observatory} (henceforth {\it Swift}), a multiwavelength observatory, has made great contributions to the understanding of the GRB phenomenon since its launch in 2004~\citep{2004ApJ...611.1005G}. The Burst Alert Telescope \citep[BAT;][]{2005SSRv..120..143B} on board {\it Swift} triggers during the prompt emission of a GRB and broadcasts the location to the ground stations.
The rapid slewing capacity of {\it Swift} enables the two onboard narrow-band telescopes, the X-ray telescope (XRT; ~\citet{2005SSRv..120..165B}), and the Ultraviolet Optical Telescope (UVOT; ~\citet{2005SSRv..120...95R}) to observe the X-ray and optical/UV emission within $\sim$1 min with observations continuing up to several days after the trigger. X-ray emission has even been observed in one case for several years after the trigger
\citep{2016MNRAS.462.1111D}. One of the most important discoveries by {\it Swift}/XRT is the existence of a canonical X-ray light curve with five power law components~\citep{2006ApJ...642..354Z}: a steep decay, plateau, normal decay, post jet-break decay, and an X-ray flare (in ~50\% of cases; \citet{2007ApJ...671.1921F}).

In addition to UVOT, early IR/optical follow-up is also possible with ground-based robotic telescopes such as MASTER-net~\citep{2010AdAst2010E..30L} and BOOTES~\citep{1999A&AS..138..583C}. The MASTER-net (Mobile Astronomical System of Telescope-Robots) includes  eight observatories located in Russian, South Africa, Spain (Canarias), and Argentina: MASTER-Amur, MASTER-Tunka, MASTER-Ural, MASTER-Kislovodsk, MASTER-Tavrida, MASTER-SAAO, MASTER-IAC, and MASTER-OAFA. MASTER-net began operating in full mode in 2010~\citep{2004AN....325..580L,2010AdAst2010E..30L,2012ExA....33..173K, 2013ARep...57..233G}. Each MASTER-II telescope contains a twin-tube aperture system with a total field of view of 8 square degrees with a photometer in the Johnson-Cousins system and polarising filters that were manufactured using linear conducting nanostructure technology~\citep{2012ExA....33..173K,2013ARep...57..233G,2012ExA....33..173K,0957-4484-16-9-076}.

The Burst Observer and Optical Transient Exploring System (BOOTES\footnote{\url{http://bootes.iaa.es}}) has been part of the effort to follow-up GRBs since 1998~\citep{1999A&AS..138..583C,2012ASInC...7..313C}. Each BOOTES station has a Ritchey-Chretien 60 cm aperture fast-slewing telescope, which cover a 10'$\times$10' field of view and is equipped with clear, Sloan g r i, and WFCAM/VISTA Z and Y filters. Each system operates autonomously. Swapping from a pre-planned target list to active observations of GRBs is accomplished by switching the filters, focussing, and pointing the telescope to the event coordinates received from the Gamma-ray Coordinates Network \citep[GCN;][]{1998AAS...192.4311B}. Thanks to the capability to react autonomously and to slew promptly, the robotic telescope has increased optical samples, particularly during the early epoch immediately following a GRB trigger. 

In this paper, we present observations of GRB 140629A, a GRB that was observed from $\sim$40 s after the trigger \citep{2014GCN.16478....1Y} as a consequence of the rapid dissemination of alerts from space to ground-based telescopes. This allowed us to obtain rich multiwavelength data from early to late epochs ($\sim$ 4 d), making this object a good case study for constraining jet properties and host environment. We present multiwavelength observations performed by $\it Swift$, $Konus$-WIND, $\it Spitzer,$ and various ground-based facilities worldwide as well as results of our modelling of the jet and its properties. The rest of the paper is organised as follows: the observations and data reduction are presented in Section 2, the analysis of the afterglow and its host galaxy are given in Section 3; and discussions are presented in Section 4. The concordance cosmology adopted in our analysis has parameters of $H_{0}$ = 71 kms$^{-1}$Mpc$^{-1}$,$\Omega_{M}$ = 0.3 and $\Omega_{\Lambda}$ = 0.7. Errors are given at 1$\sigma$ unless otherwise stated.

%-------------------------------------------------------------------------------------------------------------
%%%%%%%%%%%%%%%%%%%%%%%%%%%%%%%%%%%%%%%%%%%%%%
\section{Observations and data reduction}
\label{Observations}
\subsection{High-energy observations}
The {\it Swift}/BAT triggered and located GRB 140629A~on June 29, 2014 at 14:17:30 UT (T$_0$) \citep{2014GCN.16477....1L,2009MNRAS.397.1177E}. The BAT light curve is multiply-peaked with a duration \footnote{T$_{90}$ is burst duration defined as the time interval over which 5\% to 95\% of the counts are accumulated.} T$_{90}$=42$\pm$14.3\,s (see Figure~\ref{all LC}) and exhibited a peak count rate of $\sim$ 2000 counts/s in the 15-350 keV range at $\sim$ 0\,s after the trigger. The time-averaged spectrum from T$_{0}$-7.53 to 56.47\,s was fitted by a simple power law model with a photon index 1.86 $\pm$ 0.11~\citep{2014GCN.16481....1C}. The prompt emission light curve from BAT is shown in Figure~\ref{all LC}. GRB 140629A triggered the S2 detector of the {\it Konus}-WIND GRB spectrometer at 14:17:30:00 UT in waiting mode~\citep{2014GCN.16495....1G}. This instrument observed a double-peaked light curve~\footnote{\url{http://www.ioffe.rssi.ru/LEA/GRBs/GRB140629A/}}. A power law with an exponential cut-off is the best fit model to the time integrated spectrum with parameters $\alpha$ = $1.42\pm0.54$ and $E_{p}$ = $86\pm17$ keV. The spectrum resulted in a fluence of $3.4(\pm0.5)\times 10^{-6}$ erg/cm$^{2}$ in the 20-10000 keV energy range. The isotropic energy release in rest frame is $E_{r,iso}=4.4\times10^{52}$erg~\citep{2014GCN.16495....1G}.

The {\it Swift}/XRT began observing the field 94.2\,s after the BAT trigger and found a bright, fading uncatalogued X-ray source. An astrometrically corrected X-ray position was reported of RA(J2000)=$16^{h}$ $35^{m}$ $54.52^{s}$, Dec(J2000)=+$41^{\circ}$ $52'36.8''$ with an uncertainty of $1.7''$~\citep[90\% confidence radius;][]{2014GCN.16479....1E}. The initial XRT spectral analysis resulted in a power law photon index of $1.98\pm0.10$ and a column density of 5.2 $(+2.2, -2.0)$ $\times$ $10^{20}$ cm$^{-2}$ (90 $\%$ confidence), in excess of the galactic value at 3.5$\sigma$~\citep[9.3 $\times$ $10^{19}$ cm$^{-2}$;][]{2014GCN.16490....1O}. 

\subsection{Optical observations}
\subsubsection{MASTER}
Three stations of MASTER-net observed GRB 140629A: MASTER-Amur (in Blagoveschensk), MASTER-Tunka (near Baykal Lake), and MASTER-Kislovodsk~\citep{2014GCN.16478....1Y,2014GCN.16507....1G}.The MASTER II robotic telescope in Blagoveschensk pointed to GRB 140629A 33 s after the BAT trigger time (T$_0$) and 15 s after notice time at 14:18:03.19 UT, June 29~\citep{2014GCN.16478....1Y} and was the first ground-based telescope to observe the burst. The first two MASTER observations were obtained during the gamma-ray emission. A transient object of brightness 14.26$\pm0.06$ mag was detected. Unfortunately, observations at Blagoveschensk were carried out in only one of the two tubes of the twin-tube aperture system as a result of technical disrepair. Observations at this location lasted until $\sim$800\,s after the trigger and finished when weather conditions became unsuitable. During this time ten images with increasing exposure from 10 s to 120 s were obtained.

The MASTER II robotic telescope in Tunka pointed to GRB 140629A 78 s after T$_0$ on June 29, 2014,14:18:48.10 UT, during the evening twilight sky (the Sun was about five degrees below the horizon). For this reason, the first few images are overexposed. Nevertheless, the object is visible at the 4$\sigma$ level in one polarisation at 14:36:16\,UT (1060 s after trigger) with 3 min exposure during the evening sky observations. Following this, a small pause in observations was made for focussing. The observations were restarted in Tunka at 15:01:25 in the R and V filters. From 15:31:52 ($\sim$2600\,s after trigger) observations were performed with two mutually perpendicular polarisers. Observations continued until dawn at 18:51:36\,UT ($\sim$4.5\,h after trigger).

MASTER II in Kislovodsk pointed to GRB~140629A approximately 3.2\,h after T$_0$, which was when the weather conditions first became suitable after sunset. A total of about 40 good frames each of 180\,s exposure were obtained in white light (C) and R filters. The optical transient is not detected in individual images but is visible in summed images. Frames were grouped into three sets, added together, and processed.

\subsubsection{{\it Swift}/UVOT}
Following the detection by {\it Swift}/BAT and XRT, the UVOT began settled observations 101.15\,s after the BAT trigger and detected a fading candidate consistent with the XRT error circle~\citep{2014GCN.16494....1B}. A series of images was taken with $v$, $b$, $u$, $uvw1$, $uvm2$, $uvw2,$ and $white$ filters. The source was detected in all filters, except $uvm2$ and $uvw2$. 

\subsubsection{BOOTES}
The 60\,cm robotic telescope BOOTES-2/TELMA, in La Mayora, Malaga, Spain~\citep{2012ASInC...7..313C}
automatically responded to the GRB alert as soon as its position was observable. Observations started on June 29,
22:19:47.227 UT, $\sim$8 hrs after the {\it Swift}/BAT trigger, in the clear and Sloan-i band filters, with exposure of 60 s. The source was observed until 2014-06-30 03:46:32.804 UT, $\sim$13.5\,hrs after the burst. The clear exposures were smeared and were thus discarded. For the i-band exposures the object was faint and not visible in the single frames, but it was detectable in stacked images.

\subsubsection{OSN}
At the Sierra Nevada mountain range (Granada, Spain), the 1.5 m telescope of Sierra Nevada Observatory (OSN)\footnote{\url{http://www.osn.iaa.es/}} pointed to the source at 2014-06-29 21:06:27.23\,UT  $\sim$6.82\,hrs after trigger. The GRB field was also observed on June 30 and July 3. A series of images were obtained in Johnson-Cousins broadband filters: R filter with 300\,s exposure and V, I filters with 600 s exposure.

\subsubsection{BTA}
The optical counterpart of GRB 140629A was also observed with the 6 m Big Telescope Alt-azimuth (BTA) of SAO-RAS (Caucasus Mountains, Russia) on June 29, starting 4.1 h after the detection of the burst by {\it Swift}~\citep{2014GCN.16477....1L,2014GCN.16478....1Y}. Observations of the field were carried out with the Scorpio-I optical reducer~\citep{2005AstL...31..194A} set in the BTA primary focus. Long-slit spectroscopy was also taken with the grism VPHG440, covering 4000 to 9800 \AA. A 43.6 min spectral observation was obtained. The particular configuration of the device in combination with the $1^{\prime\prime}$ slit achieves a resolution of full width at half maximum (FWHM)=13 \AA. 

\subsubsection{GTC}
The 10.4 m Gran Telescopio CANARIAS\footnote{\url{http://www.gtc.iac.es}} (GTC, Canary Islands, Spain) obtained several images with the Optical System for Imaging and low Resolution Integrated Spectroscopy (OSIRIS) camera ~\citep{2000SPIE.4008..623C} on Feburary 27, 2015 and February 7, 2017, $\sim$8 months and 2.7\,yr after the burst respectively in order to detect the host galaxy. Eight images were obtained in the first epoch with Sloan-g, r, i filters. Four images were taken with the Sloan-i filters of 90\,s exposure, three images were taken with the Sloan-g filter of 140\,s exposure, and one 90\,s exposure was taken with the Sloan-r filter. In the second epoch, 22 images were obtained: seven images each in the Sloan-g and Sloan-r filters with 150\,s exposure and 120\,s exposure, respectively, and eight images with 90\,s exposure in Sloan-i band.

\subsection{Infrared observations}
The {\it Spitzer} Space Telescope (SST) also observed the source with the InfraRed Array Camera (IRAC) instrument at a wavelength of 3.6 $\mu m$ \citep{2016ApJ...817....7P,2016ApJ...817....8P}. The total exposure time is 2\,h. The data were downloaded from the {\it Spitzer} data archive center\footnote{\url{http://sha.ipac.caltech.edu/applications/Spitzer/SHA/}}. The source was observed on June 6, 2015, $\sim$1\,yr after the trigger. 

\section{Data analysis and results}
\subsection{Photometry}
The final photometry for the MASTER telescopes was extracted using the \textsc{IRAF\footnote{IRAF is distributed by the National Optical Astronomy Observatory, which is operated by the Association of Universities for Research in Astronomy, Inc. under cooperative agreement with the National Science Foundation. http://ast.noao.edu/data/software}} package~\citep{1993ASPC...52..173T}. The
MASTER observations were taken with the polariser R, V, and C bands (P0, P45, P90, V, R, C at Fig.1). The C filter is white light corresponding to 0.2B+0.8R. The polarisation observations were taken with orientations 0$^\circ$, 45$^\circ$, and 90$^\circ$ to the celestial equator. Automatic astrometric and photometric calibrations were performed with a method common to all MASTER observatories~\citep{2012ExA....33..173K,2013ARep...57..233G}. For these data, a robust \textquoteleft centroid\textquoteright\  algorithm was used to determine the background level. This algorithm allows us to exclude the influence of nearby objects. The data were corrected for the fluctuations with atmospheric opacity using the  Astrokit\  programme~\citep{2014AstBu..69..368B}, which implements a slightly modified algorithm to that described in~\citet{2001PASP..113.1428E}. This programme conducts differential photometry using an ensemble of stars that are close to an object. The details of the photometry calibration can be found in~\citet{2012MNRAS.421.1874G}. For the polarisation observations, stars with zero polarisation are required for the channel calibration. We assume the polarisation of light from stars in the field of view is small. This can be checked using Serkowski law~\citep{1975ApJ...196..261S}. The difference in magnitudes between two polariser orientations averaged for all reference stars gives the correction that takes into account different channel responses.

{\it Swift}/UVOT sky images were downloaded from the {\it Swift} science data centre\footnote{\url{http://www.swift.ac.uk}} and the magnitudes
were extracted following standard UVOT procedure~\citep{2008MNRAS.383..627P}. In this work, a 3$\sigma$ upper limit is given when signal to noise
is <3. For individual filters after 2000 s, the data are binned with $(\delta t)/t$ = 0.2 to improve the signal to noise.

In order to obtain the instrumental magnitudes for the other instruments, point spread function (PSF) photometry was applied with the DAOPHOT
tool in the IRAF package. Photometric magnitudes from the OSN were calibrated with the nearby reference stars in USNO-B1, GSC2.3 catalogue~\citep{2003AJ....125..984M,2008AJ....136..735L}. For
the GTC, magnitudes were calibrated with the standard star $STD\_$PG1323-086D.

\begin{figure}
  \includegraphics[width=17cm,angle=0,scale=.57]{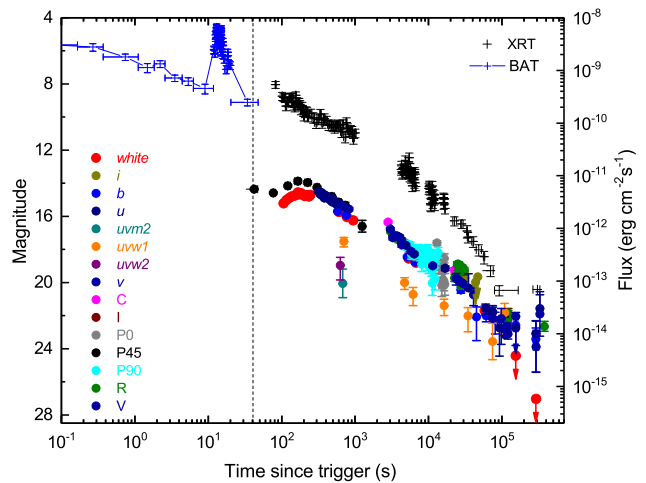}
 \caption{X-ray and optical light curves for GRB 140629A. Observations from both BAT and XRT are given with blue crosses and black crosses, respectively. The BAT data are normalised to the same energy range as XRT. The optical data are shown with circles. The vertical dash line indicates the end of the prompt emission, given by $T_{90}$.}
 \label{all LC}
 \vspace{-0.4cm}
\end{figure}

The observation log of GRB 140629A is given in Table~\ref{Tablelog} and the photometry for all filters and polarisations is presented in Table~\ref{Table6}. All magnitudes are presented in Vega system except the GTC host galaxy observations, which are calibrated using the AB system. The final magnitude errors include the systematic error from the reference star calibration. The magnitudes in the table are not corrected for galactic extinction owing to the reddening of E(B-V) = 0.01 in the direction of the burst~\citep{1998ApJ...500..525S}. For clarity, the afterglow light curves are shown in Figure~\ref{all LC}.

\begin{table}
        \caption{Observation log of GRB 140629A.}
        \label{Tablelog}
        \centering
        \begin{tabular}{c c c}
                \hline
                  $t_{start}$ & $t_{end}$ & Filters\\
                \hline
                \multicolumn{3}{c}{MASTER} \\           
                2014 Jun 29 14:18:03&Jun 29 21:58:09&C,V,R\\
                \multicolumn{3}{c}{$\it{Swift}$/UVOT} \\ 
                ~&&\textit{white,u,v,b}\\
                2014 Jun 29 14:19:10&Jul 03 09:33:28&\textit{uvw1,uvw2}\\
                ~&&\textit{uvm2}\\
                \multicolumn{3}{c}{BOOTES} \\
                2014 Jun 29 22:19:47& Jun 30 03:47:33&\textit{i}\\
        \multicolumn{3}{c}{OSN} \\ 
                2014 Jun 29 21:06:27&Jun 29 22:54:33&V,I,R\\
                2014 Jun 30 22:20:55&Jul 01 00:19:09&V,I,R\\
                2014 Jul 03 22:52:42&Jul 04 00:16:30&R\\
                \multicolumn{3}{c}{GTC} \\ 
                2015 Feb 27 06:03:20&Feb 27 06:23:50&\textit{u,g,r,i}\\
                2017 Feb 06 04:43:04&Feb 06 05:36:08&\textit{u,g,r,i}\\
                \multicolumn{3}{c}{\it Spitzer} \\
                2015 Jun 05 16:48:20&Jun 05 19:00:48&3.6\textit{um}\\
                \hline
                \vspace{-0.7cm}
        \end{tabular}
\end{table}

\subsection{Temporal properties of the afterglow: An empirical fit}
In this section, we fitted the light curves in X-ray and optical band with the empirical multi-segment smooth broken power law models ~\citep{1999A&A...352L..26B,2006ApJ...640L...5J,2007A&A...469L..13M}. 

\begin{figure}
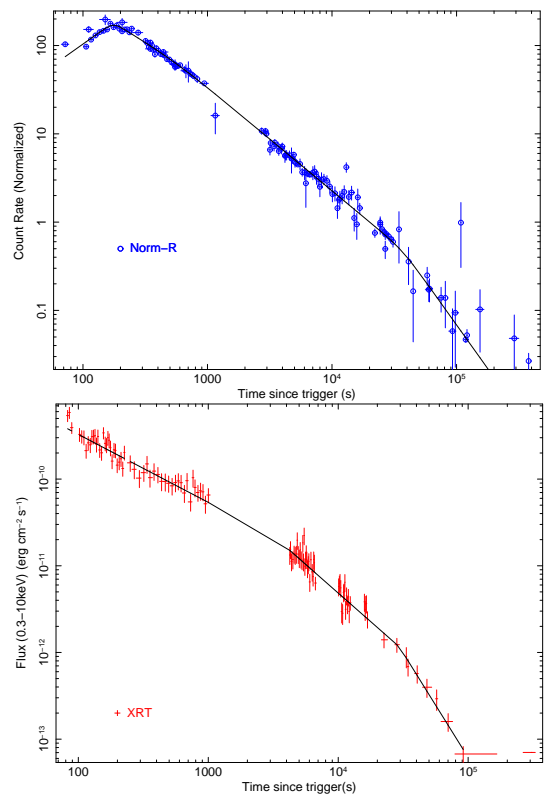

        \includegraphics[angle=270,scale=.3]{optical.ps}
        \includegraphics[angle=270,scale=.3]{xray.ps}
        \caption{Fitting of the X-ray and optical GRB 140629A data. The normalised optical data are shown on the top panel with blue circles. The X-ray data are plotted on the bottom panel with red crosses. The black lines indicate the fitting results.}
        \label{first figure}

\end{figure}

The X-ray light curve (0.3-10 keV) was obtained from the UK Swift Science Data Centre at the University of Leicester~\citep{2009MNRAS.397.1177E}. As shown in
Figure \ref{first figure}, the GRB 140629A X-ray light curve appears to show a canonical structure. An initial shallow decay is followed by a normal
decay and then a steep decay~\citep{2006ApJ...642..354Z}. To ensure two breaks are required, we first attempted to fit the light curve with a single
broken power law. This resulted in a poor fit with a reduced chi square ($\chi^{2}$)$=1.32~(110~d.o.f.)$, and corresponding null hypothesis probability of only $0.01\%$.
We then tested a smooth broken power law, which showed an improvement giving a reduced $\chi^{2}=1.10~(110~d.o.f.)$. For this model the best fitting
parameters are $\alpha_{1}=0.84\pm0.02$, $\alpha_{2}=1.87\pm0.08$ with a break time $(8.8\pm1.3)\times10^{3}$ s. We then tried a smooth double broken
power law model. This again improved the fit giving a reduced $\chi^{2}/d.o.f.$= 0.99/108. According to the Akaike information
criterion~\citep[AIC]{1974ITAC...19..716A,2007MNRAS.377L..74L}, the smooth double broken power law model gives a lower AIC value in comparison to the
smooth broken power law, suggesting the second break is required. This is also confirmed by the F-test, which suggests the second break is statistically required at more than 3$\sigma$ confidence. We can also check the need for an additional break using a Monte Carlo simulation. We create 10000 synthetic light curves by randomly selecting, for each data point of the observed light curve, a new flux and flux error using a Gaussian function for which the mean and standard deviation is equal to the original observed flux and flux error. Each of the synthetic light curves are then fitted with both a broken power law and a double broken power law. From the resulting distribution of the change in reduced $\chi^{2}$, we find that 98.2\% of the simulated light curves have a change in reduced $\chi^{2}$ that is equal to or greater than that obtained for the observed X-ray light curve. The Monte Carlo simulation thus suggests that the double break power law is preferred over the broken power law model at the 2$\sigma$ confidence level. We do not identify any X-ray flares in the light curve.

For the optical data, we normalised the observations in the different filters to Johnson-R band using the period between 3000 s and 30000 s, during
which the light curves appear to decay in the same fashion~\citep{2007MNRAS.380..270O}. The resulting light curve is shown in Figure \ref{first figure}.
We exclude data prior to 70\,s from our analysis since the data are likely contaminated by the prompt emission. When fitting the optical data, we tested both
a smooth double broken and a smooth triple broken power law against the data. The reduced $\chi^2$ changed from 1.53 (d.o.f = 123) to 1.38 (d.o.f=121)
with the addition of third break. The smooth triple broken power law model is preferred according to F-test, which provides a chance probability of 0.0008.
We also used the Monte Carlo method, which we used to determine the significance of improvement of an additional break in the X-ray light curve, on the optical data. The synthetic light curves in optical are fitted with both a double broken and a triple broken power law model. In the resulting distribution of the change in reduced $\chi^{2}$, we find that 99.9\% of the simulated light curves have a change in reduced $\chi^{2}$ that is equal or greater than that obtained from the observed light curve. Thus the Monte Carlo simulation suggests the triple break power law is preferred over the double break power law at 3$\sigma$ confidence level. The values of the best fitting parameters for the optical and X-ray light curves are shown in Table \ref{Table1} and the temporal fits are shown in Figure~\ref{first figure}. 

        \vspace{-0.5cm}
%%%%%%%%%%%%%%%%%%%%%%%%%%%%%%%%%
\begin{table}
        \caption{Results of the best fit model to the X-ray and optical afterglows of GRB 140629A. }
        \label{Table1}
        \centering
        \begin{tabular}{ c c c c}
                \hline
                \multicolumn{2}{c}{Optical}&\multicolumn{2}{c}{X-ray}\\
                \hline
                Para & Value\tablefootmark{a} & Para & Value\tablefootmark{a}\\
                \hline
                $\alpha_{o,1}$ & $-0.72^{+0.15}_{-0.33}$ &  &  \\
                $t_{o,b_1}$ & $176.85^{+3.48}_{-3.22}$ &  & \\
                $\alpha_{o,2}$ & $0.91^{+0.03}_{-0.04}$ & $\alpha_{x,1}$ & $0.78^{+0.04}_{-0.04}$ \\
                $t_{o,b_2}$ & $638.69^{+126.31}_{-105.89}$ & $t_{x,b_1}$ & $3428.52^{+1167.48}_{-808.52}$ \\
                $\alpha_{o,3}$ & $1.17^{+0.01}_{-0.01}$ & $\alpha_{x,2}$ & $1.33^{+0.09}_{-0.07}$ \\
                $t_{o,b_3}$ & $36164.96^{+7895.06}_{-5064.96}$ & $t_{x,b_2}$ & $31179.38^{+12470.62}_{-6560.38}$ \\
                $\alpha_{o,4}$ & $1.97^{+0.18}_{-0.10}$ & $\alpha_{x,3}$ & $2.46^{+0.49}_{-0.24}$ \\
                $\chi^{2}/d.o.f.$ & 1.38/121 &$\chi^{2}/d.o.f.$ & 0.99/108 \\
                \hline
                \vspace{-0.4cm}
        \end{tabular}
        \tablefoottext{a}{\small The break times are given in seconds.}
\end{table}

\subsection{Spectral analysis}
\subsubsection{Optical spectroscopy}
\begin{figure*}
        \centering
        \includegraphics[width=17cm,angle=0,scale=0.65]{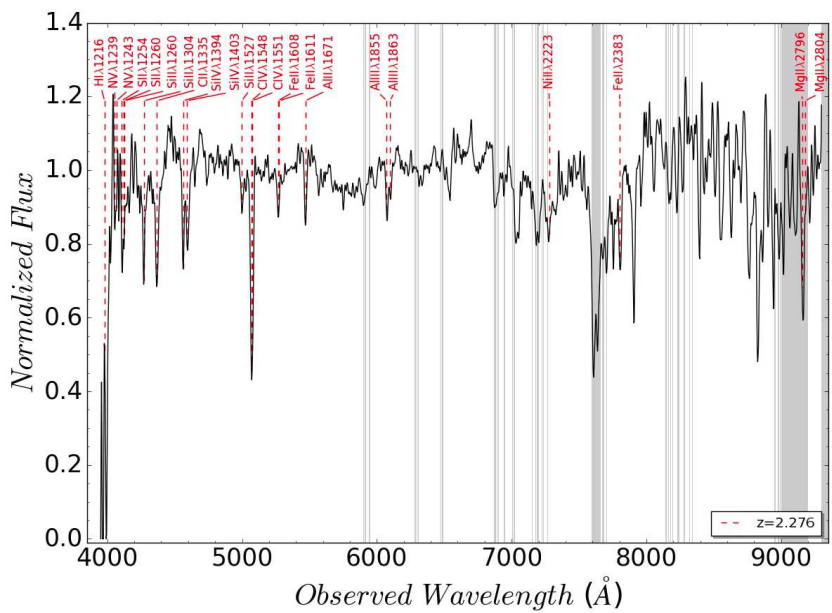}
        \caption{ Overall view of the optical spectrum from BTA, obtained $\sim$4.1 hr after the GRB 140629A trigger. These metal lines of the
          absorption system in the GRB host galaxy are labelled in red, showing the corresponding transitions. The wavelength range with strong
          telluric absorption features are indicated by grey vertical lines.
        }
        \label{Spectroscopy}
\end{figure*}

The optical spectrum observed by the 6 m BTA telescope, given in Figure \ref{Spectroscopy}, shows multiple absorption lines that we identify
as Lyman-$\alpha$ (Ly-$\alpha$) absorption, 
\ion{Al}{III} (1854.72\AA, 1862.78\AA), 
\ion{C}{IV} (1548.20\AA, 1550.77\AA), 
\ion{C}{II} (1334.53\AA), 
\ion{N}{V} (1238.81\AA, 1242.80\AA)
\ion{Fe}{II} (1608.45\AA), 
\ion{Mg}{II} (2803.53\AA, 2796.35\AA), 
\ion{Si}{ii} (1260.42\AA, 1304\AA, 1526.72\AA),
\ion{Si}{IV} (1393.76\AA, 1402.77\AA), and
\ion{Al}{II} (1670.79\AA).
All these absorption features can be attributed to a single intergalactic cloud at a common redshift z= 2.276 $\pm$0.001. This measurement is consistent with and refines previous determinations~\citep{2014GCN.16489....1M,2014GCN.16493....1D,2018ApJ...860....8X} of the redshift of the GRB and its host galaxy. Both random and systematic errors are included in the uncertainty of the redshift. There are a few absorption features probably due to intervening systems in the line of sight, but we are not able to identify nor determine their redshift. In addition, we do not find any obvious strong emission lines in our spectrum. The absorption lines associated with the host galaxy at z=2.276 are identified on the
spectrum provided in Fig \ref{Spectroscopy}.

We measured the equivalent widths (EW) of the detected absorption features (see Table \ref{TableEW}). We found that the \ion{C}{IV} line has a rest-frame EW value of 4.11$\AA$. This makes it the strongest absorption feature in the spectrum, confirming the identifications made
by \citet{2012A&A...548A..11D} and \citet{2018ApJ...860....8X}. The EWs of the high ionisation species are higher than average, as compared to the
results of \citet{2012A&A...548A..11D}, while the low ionisation species show no peculiarities. This implies the line of sight has a stronger
ionisation absorption than is typically found for GRBs. We are also able to derive the EW ratio of $\ion{C}{IV}/\ion{C}{II}=2.72\pm0.15$,
which is consistent with the result of \citet{2018ApJ...860....8X}, but higher than the median value found for GRBs in \citet{2012A&A...548A..11D}. 
We find the ratio of $\ion{Al}{III}/\ion{Al}{II}=1.44\pm0.09$ and the ratio of $\ion{Si}{IV}/\ion{Si}{II}= 5.1$, which are both higher than the median
values found for GRBs \citet{2012A&A...548A..11D}. Using the ratios of $\ion{C}{IV}/\ion{C}{II}$ and $\ion{Si}{IV}/\ion{Si}{II}$, we find this GRB
to be consistent with the highly ionised tail of the distribution of ionisation ratios of carbon and silicon; see Fig.11 in~\citet{2012A&A...548A..11D}.

\begin{table}
        \caption{Spectroscopic information for the GRB 140629A. }
        \label{TableEW}
        \centering
        \begin{tabular}{p{22pt} p{46pt} c p{25pt} }
                \hline
                Wave & Rest EW& Feature & $z$ \\
                ({\AA}) & ({\AA}) & & \\
                \hline
                4057.8 & 0.95 $\pm$ 0.12 & \ion{N}{V}{$\lambda$}1238.8 &  2.27553 \\
                4071.2 & 0.51 $\pm$ 0.07 & \ion{N}{V}{$\lambda$}1242.8 &  2.27581 \\
                4127.5 & $<$ 0.48  &\ion{S}{II}{$\lambda$}1259.52 + \ion{Si}{II}{$\lambda$}1260.42 & -- \\
                4272.5 & 2.01 $\pm$ 0.11 &\ion{O}{i}{$\lambda$}1302.170 + \ion{Si}{II}{$\lambda$}1304.4 & -- \\
                4289.5 & 0.66 $\pm$ 0.06 & \ion{Si}{II}{$\lambda$}1309.3 & 2.27622 \\
                4374.9 & 1.51 $\pm$ 0.08 & \ion{C}{II}{$\lambda$}1334.5 + \ion{C}{II}*{$\lambda$}1335.7 & -- \\
                4584.6 & 2.96 $\pm$ 0.11 & \ion{Si}{IV}{$\lambda$}1393.8+1402.8 & -- \\
                5007.2 & 0.58 $\pm$ 0.08 & \ion{Si}{II}{$\lambda$}1526.71 & -- \\
                5077.4 & 4.11 $\pm$ 0.07 & \ion{C}{IV}{$\lambda$}1548.2+1550.8 & -- \\ 
                5275.9 & 0.84 $\pm$ 0.06 & \ion{Fe}{II}{$\lambda$}1608.4+1611.2 &  -- \\
                5474.7 & 1.01 $\pm$ 0.05 & \ion{Al}{II}{$\lambda$}1670.8 & 2.27672 \\
                6085.6 & 1.45 $\pm$ 0.06 & \ion{Al}{III}{$\lambda$}1854.7+1862.8 & -- \\
                7809.7 & 1.59 $\pm$ 0.16 & \ion{Fe}{II}{$\lambda$}2383.8 & 2.27616 \\
                9161.8 & 2.66 $\pm$ 0.13 & \ion{Mg}{II}{$\lambda$}2796.4+2803.5 & -- \\
                \hline
        \end{tabular}
\end{table}

\subsubsection{Afterglow spectral analysis}
As reported above, our temporal analysis of the X-ray light curve shows it to be best fit by a double broken power law that has two breaks at $\sim$3000 s and $\sim$30000 s. 
For each X-ray segment, we extracted an X-ray spectrum from the {\it Swift}/XRT GRB spectrum repository\footnote{\url{http://www.swift.ac.uk/xrt_spectra/}} \citep{2009MNRAS.397.1177E}. We fit each spectra with a power law and two photoelectric absorption components: one for our Galaxy and the other for the host galaxy of the burst. The fitting results are shown in Table \ref{Table2}. The spectral indices of the three spectra are consistent with each other at 1$\sigma$ confidence. Therefore, the spectral slope does not show any evidence for evolution
across the three X-ray light curve segments. 

\begin{figure}
        \includegraphics[width=17cm,angle=270,scale=.35]{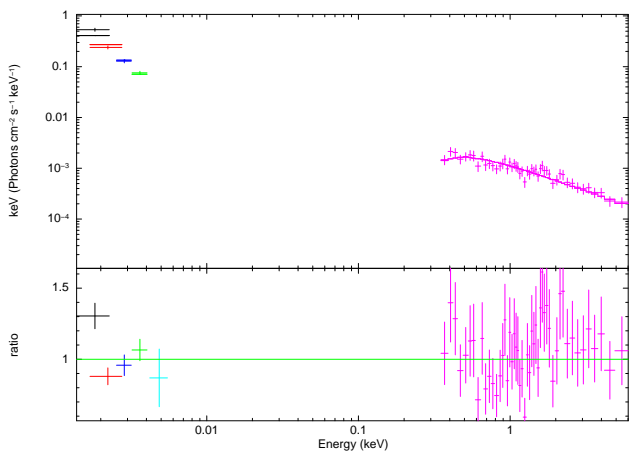}
        \caption{Optical and X-ray SED in time interval of 9350s fitted with the SMCxBKP model.}
        \label{spectrum}
\end{figure}

In order to constrain the spectral properties of the optical and the X-ray afterglow, we produced spectral energy distributions (SEDs) at $\sim$775 s and $\sim$9350 s after trigger, respectively. The joint SEDs (see Figure \ref{spectrum}) were fit using Xspec 12.9.0 in the HEAsoft package~\citep{1996ASPC..101...17A}. We fit both a power law and a broken power law model to each of the SEDs, including components for dust and photoelectric absorption for both our Galaxy and the GRB host galaxy. We expect the synchrotron cooling frequency to be the cause of the spectral break in the broken power law model, therefore we fixed the difference in the two spectral indices to be $\Delta\beta=$0.5.
For the extinction in our Galaxy, we fixed the dust component to have an $E_{B-V}=0.0067$~\citep{1998ApJ...500..525S} and used the Milky Way (MW) extinction curve. We fixed the hydrogen column density of the MW to be 9.3$\times$ $10^{19}$cm$^{-2}$~\citep{2005A&A...440..775K,2014GCN.16490....1O}. 
The fitting results are listed in Table \ref{Table3}. We tested three extinction laws for the GRB host galaxy: $R_{V}$=3.08, $R_{V}$=2.93 and $R_{V}$=3.16 for the MW, Small Magellanic Cloud (SMC), and Large Magellanic Cloud (LMC), respectively.

In the SED obtained at 775s, we consider the power law to be the best model, since the F-test indicates that the broken power law model does not provide a significant improvement. Of the extinction models, the MW model provides the best chi-square of the three extinction models, but the reduced $\chi^2$ is similar for all three scenarios. For the SED at 9350s (see Fig \ref{spectrum}), we find that the SMC model gives a better fit compared to the other two extinction models for both the power law and broken power law models. Again for this SED, the F-test indicates that the broken power law model does not improve the fit compared to the single power law model. The AIC supports this conclusion, as the AIC value increases for the broken power law model compared to the power law model. While there is no strong preference for a particular extinction law for the 775 s, the 9350 s SED clearly indicates the SMC extinction law is the best model. This is consistent with the preference for an SMC extinction law found for a large number of GRBs \citep{2010MNRAS.401.2773S}. We therefore assume this model during further investigation of this GRB. Comparing the SMC power law models of both SEDs, we find the parameters from the 775 s and 9350 s SEDs are consistent at 3$\sigma$ confidence level and that the $N_{H}$ values are consistent at 1$\sigma$ confidence with the best fit values determined from the X-ray spectrum.

\begin{table}
        \caption{Spectral analysis of the X-ray light curve of the GRB 140629A afterglow fitted with
                three segments.}
        \label{Table2}
        \centering
        \begin{tabular}{c c c c }
                \hline
                Segment & Time interval  & Photon&$N_{H}$\\
                &  (s) & index&($10^{21}$cm$^{-2}$)\\
                \hline
                1&  $100-3\times10^{3}$ &$1.86_{-0.13}^{+0.14}$&$6.9_{-4.0}^{+4.4}$\\
                2&  $3\times10^{3}-3\times10^{4}$ &$1.93_{-0.11}^{+0.11}$&$7.1_{-3.3}^{+3.6}$\\
                3&  $3\times10^{4}-10^{5}$ &$1.91_{-0.33}^{+0.36}$&$7.0_{-7.0}^{+12.5}$\\
                \hline
        \end{tabular}
\end{table}

\begin{table}[ht]
        \caption{Fit results for the GRB 140629A afterglow SEDs. }
        \label{Table3}
        \centering
        \begin{tabular}{c p{25pt} c p{30pt} c c}
                \hline
                Model\tablefoottext{a}& R-chi.& $E_{B-V}$ & $N_{H}$  & Photon\\
                & /D.O.F & (mag)  & ($10^{21}$cm$^{-2}$) & index \\
                \hline
                775s&&&&\\
                \hline
                MW$\times$POW& 0.94/27 & $0.131^{+0.017}_{-0.017}$ &$3.30^{+2.24}_{-2.02}$ &$1.948^{+0.026}_{-0.026}$\\
                LMC$\times$POW&1.02/27&$0.108^{+0.013}_{-0.013}$&$3.69^{+2.27}_{-2.04}$&$1.963^{+0.027}_{-0.027}$\\
                SMC$\times$POW&1.05/27&$0.085^{+0.011}_{-0.011}$&$3.40^{+2.24}_{-2.02}$&$1.952^{+0.025}_{-0.026}$\\
                MW$\times$BKP&0.94/26&$0.160^{+0.006}_{-0.006}$&$6.23^{+2.13}_{-1.90}$&$1.563^{+0.010}_{-0.010}$\\
                LMC$\times$BKP&0.99/26&$0.125^{+0.005}_{-0.005}$&$4.67^{+2.05}_{-1.84}$&$1.502^{+0.010}_{-0.010}$\\
                SMC$\times$BKP&1.02/26&$0.098^{+0.004}_{-0.004}$&$4.35^{+2.04}_{-1.83}$&$1.489^{+0.010}_{-0.010}$\\
                \hline
                9350s&&&&\\
                \hline
                MW$\times$POW&1.92/50 & $0.135^{+0.017}_{-0.017}$ & $6.52^{+2.01}_{-1.86}$ &$2.020^{+0.024}_{-0.024}$\\
                LMC$\times$POW&1.44/50&$0.122^{+0.013}_{-0.013}$&$7.51^{+2.05}_{-1.89}$&$2.052^{+0.023}_{-0.023}$\\
                SMC$\times$POW&1.25/50&$0.083^{+0.010}_{-0.009}$&$7.20^{+2.01}_{-1.85}$&$2.039^{+0.020}_{-0.020}$\\
                MW$\times$BKP&1.95/49&$0.135^{+0.017}_{-0.017}$&$6.52^{+2.01}_{-1.86}$&$2.020^{+0.024}_{-0.024}$\\
                LMC$\times$BKP&1.47/49&$0.122^{+0.013}_{-0.013}$&$7.51^{+2.05}_{-1.89}$&$2.053^{+0.023}_{-0.023}$\\
                SMC$\times$BKP&1.28/49&$0.085^{+0.010}_{-0.009}$&$7.16^{+2.01}_{-1.85}$&$2.040^{+0.020}_{-0.020}$\\
                \hline
        \end{tabular}
        \tablefoottext{a}{MW is Milky Way extinction model.  LMC is Large Magellanic
                Cloud extinction model; SMC is Small Magellanic Cloud extinction model; POW is
                power law model; and BKP is broken power law model.}
\end{table}

\subsection{Host galaxy SED fitting}
\begin{figure}
        \centering
        \includegraphics[width=17cm,angle=0,scale=.4]{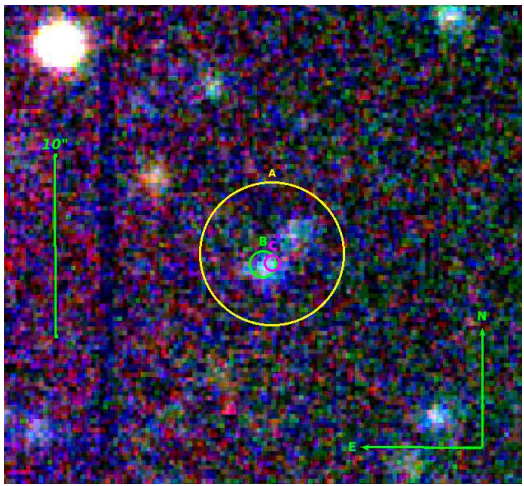}
        \caption{Sloan gri-bands false colour image of the field of GRB 140629A taken with
                the 10.4 m GTC on July 2, 2017. Circle A (yellow dash circle) represents the
                4 arcsec radius error circle of the XRT. 
                The circle B (green circle) and C (pink circle) represent the UVOT observation in 0.74 arcsec and 0.42
                arcsec, respectively~\citep{2014GCN.16477....1L,2014GCN.16494....1B}. The host galaxy is clearly found
                at the burst location. North is up and east to the left.}
        \label{Host galaxy}
\end{figure}

In order to study the GRB host galaxy, late time observations were taken by 10.4 m GTC at two separate epochs. An object was found within the
XRT and UVOT error circles in the second epoch, 2.7\,yr after the trigger (as shown in Figure \ref{Host galaxy}). {\it Spitzer} also observed
the field $\sim1$ yr after the burst in the infrared in the 3.6 $\mu m$ band. By this time, the afterglow contribution is negligible. 

The brightness of the host galaxy is 24.94$\pm$0.24 mag in Sloan-r band, which is within the brightness distribution for GRB host galaxies
(see Fig. 2. in \citet{2005A&A...441..975G}). It is slightly fainter than the reference $M^{\star}_{r}$ galaxy at the same distance, where $M^{\star}_{r}$ is the r-band absolute magnitude, considering $M^{\star}_{r}$=-20.29+5log$(H_{0}/100)$ \citep{1996ApJ...464...60L} and adopting an Einstein-de Sitter Universe model where the
spectrum of the $M^{\star}_{r}$ galaxy was assumed to be a power law with an index of 2. 

\begin{figure}
        \centering
        \includegraphics[width=17cm,angle=0,scale=.45]{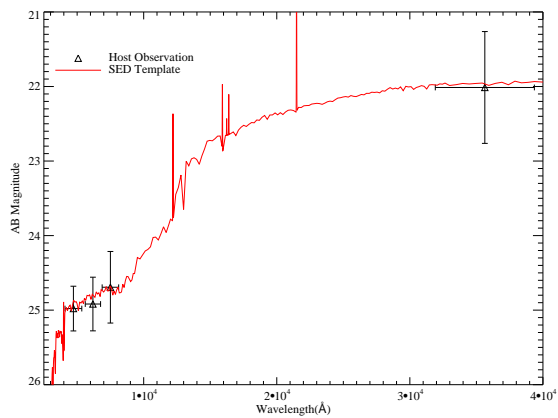}
        \caption{GRB 140629A host galaxy observations overlaid with the best fit host galaxy template.}
        \label{Host galaxy SED}
\end{figure}

The four photometric magnitudes for the host galaxy were fit with a set of galaxy templates based on the models from ~\citet{2003MNRAS.344.1000B}
at a fixed redshift ~\citep{2007A&A...475..101C,2011A&A...534A.108K} using the LePhare package (v.2.2; ~\citealt{1999MNRAS.310..540A,2006A&A...457..841I}).
As shown in Figure \ref{Host galaxy SED}, the optical SED is reproduced best (reduced $\chi^{2}$/dof=0.1/3) by a template of a galaxy with a starburst
age of $1.14_{-0.35}^{+1.03}$ Gyr and a mass of log($M_{*}/M_\odot$)=$8.3_{-0.4}^{+0.9}$ , which is lower but consistent within errors with $10^{9.3}M_\odot$,
i.e. the average value of GRB hosts~\citep{2009ApJ...691..182S}. The absolute bolometric magnitude of the host galaxy is -22.49 mag and the star formation
rate (SFR) is log(SFR)=$1.1_{-0.4}^{+0.9}$ M$_\odot$ yr$^{-1}$, which is determined from the UV luminosity of the rest-frame SED~\citep{1998ApJ...498..541K}.
The specific star formation rate (SSFR) for this burst is log(SSFR)=$-7.5_{-1.3}^{+0.6}$ yr$^{-1}$.

\subsection{Polarisation}
According to our observations with the MASTER network, GRB 140629A reached maximum optical brightness $\sim$150 s after the burst with 13.8 mag in
white light, measured using the polariser, after which it decays as a power law. The difference between the signals obtained in the two polarisers for
the time interval from 4463 s to 11596 s were computed as $Q = \frac{I_1-I_2}{I_1+I_2}$. It was found that the dimensionless time-averaged Stokes
parameter is $Q = 2.5\pm2.6\%$. For the derived uncertainty of 2.6\%, the 1$\sigma$ upper limit for the degree of linear polarisation P is about
18\% (see Fig. 14 of~\citet{2012MNRAS.421.1874G}: the value of P = 18\% matches 1$\sigma$ probability $L = 100\%-68\% = 32\%$ for the curve corresponding
to a relative accuracy 2.6\%). At the same time, a non-evolving, weak polarisation result was obtained by Hiroshima one-shot wide-field polarimeter(HOWPol) at the Kanata telescope. They
found P $\sim 2\%$  between $\sim70$ to $\sim1200$ s in the burst frame (Fig.10 in ~\citet{2016MNRAS.455.3312G}). Our upper limit is consistent
with their result. 

\section{Discussion}
We have studied the optical and X-ray afterglow of GRB 140629A. There is no strong evidence for spectral evolution, with the spectral indices consistent
within $3\sigma$. The optical light curve begins with an initial rise, which decays thereafter with two breaks. The X-ray light curve decays from the
start of observations and also decays with two breaks. A weak polarisation signal was found in the afterglow observations and we were able to obtain
information on the host by fitting the host galaxy SED. In the following subsections, we examine the closure relations between the temporal and
spectral indices. Then we use the data to test the jet structure to determine a plausible scenario to explain this burst. Finally, we explore
the properties of the host.

\subsection{Closure relationship in optical and X-ray data}
The closure relations are a set of equations that relate observational parameters, namely the spectral and temporal indices, with the microphysical parameters, for example $p$ (the electron energy spectral index), $\upsilon_{m}$ (the characteristic synchrotron frequency of the electrons at the minimum injection energy)  and $\upsilon_{c}$ (the cooling frequency). Typically, the closure relations are used to determine the location of the observing bands relative to the synchrotron frequencies, $\upsilon_{m}$ and $\upsilon_{c}$, and also the environment in which the burst occurs~\citep{1998ApJ...497L..17S,1999ApJ...520..641S,2004RvMP...76.1143P,2004IJMPA..19.2385Z,2006ApJ...642..354Z}.

For GRB 140629A, we examined the three segments of the X-ray light curve. Spectral evolution is not observed across these segments. We first examined the
second segment since this is expected to be consistent with the normal decay phase. After the first break at $\sim3000$ s, the light curve becomes
steeper with $\alpha_{x,2}=1.33^{+0.09}_{-0.07}$. This slope is typical of the normal decay phase ($\sim$1.1-1.5;~\citet{2006ApJ...642..354Z}). The spectral slope for this segment is $\beta_{X,2}=0.93^{+0.11}_{-0.11}$. During this phase, the temporal
index and spectral index are consistent with the closure relation $\alpha=3\beta/2$, which is for electrons that are slow cooling within the range
$\upsilon_{m}<\upsilon_{x}<\upsilon_{c}$ without energy injection in a uniform circumstellar medium. 
The first segment decays with $\alpha_{x,1}=0.78^{+0.04}_{-0.04}$ until 3430 s and the spectral index is $\beta_{X,1}=0.86^{+0.14}_{-0.13}$. We first tested
a simple non-injected model and found that neither $\alpha=3\beta/2$, $\alpha=(3\beta+1)/2$ and $\alpha=(1-\beta)/2$ agree with the theoretic prediction
(>3 $\sigma$). Only the relation $\alpha=(3\beta-1)/2$ in the $\upsilon_{x}>\upsilon_{c}$ case can fit the indices at ~1$\sigma$. However, comparing the
best fit closure relations between the first and second segments implies there must be a spectral break between the two segments which is not observed. We
therefore examined more complex closure relations that include energy injection. It is assumed that the luminosity evolves as $L(t)=L_{0}(t/t_{b})^{-q}$,
where $q$ is the luminosity index affected by the energy injection~\citep{2006ApJ...642..354Z}. The relation satisfied is $\alpha=(q-1)+(2+q)\beta/2$. This is for a constant density medium with slow cooling electrons, where $\upsilon_{m}<\upsilon_{x}<\upsilon_{c}$. This relation requires $q=0.59\pm0.05$.
This relation can also be used in the case $\upsilon<\upsilon_{m}$, but we can able to rule this out since the observed slope is 3$\sigma$ away from the predicted
slope. Therefore, the change between the first two segments most likely signals the end of additional energy injection, after which the afterglow enters
the normal decay phase. The origin of the shallow decay phase (plateau) is also an issue that has been debated. The plateau may be categorised as having an internal or external origin, depending on the behaviour of the temporal index of the next light curve segment. The internal plateau is followed by a steep decay whose index is larger than 3, even as large as 10. This plateau is a result of the internal dissipation of a millisecond magnetar as it spins down~\citep{2007ApJ...670..565L,2007ApJ...665..599T,2014arXiv1401.1601Y}. In this case when the energy injection ceases a sharp drop in the light curve is observed. The decay index following an external origin for the plateau is typically smaller than 3 and is well explained by energy injection into the external shocks from either slower travelling shells that are received later or by a long-lived central  engine~\citep{1998A&A...333L..87D,2001ApJ...552L..35Z,2019arXiv190507929T}. For GRB 140629A, the plateau is followed by a normal decay with a slope of 1.33, indicating that it has an external origin. The change in slope across the second break at $\sim$30000s is $\Delta\alpha\sim1.1$. We immediately ruled out several potential interpretations for this break, including the transition of the cooling frequency across the band that predicts a $\Delta\alpha\sim0.25$~\citep{1998ApJ...497L..17S}; an energy injection from refreshed shocks; a long-lasting central engine that predicts a $\Delta\alpha\sim0.7$~\citep{1998ApJ...496L...1R,2000ApJ...535L..33S,2002ApJ...566..712Z}, or an external density change, which in order to achieve such a large change in alpha, the density would have to decrease by a factor larger than $10^{3}$~\citep{2007MNRAS.380.1744N,2012ApJ...756..189F}. Therefore, this observed break can only be explained by the jet geometry, for example a jet break. The light curve after the jet break should follow $\varpropto t^{-p}$. For the third segment, the best fitting closure relation is for a spreading jet with slow cooling, where $\upsilon_{m}<\upsilon_{x}<\upsilon_{c}$ is consistent with the previous segments. Using the spectral index for X-ray segment 3, we found a temporal slope of $-2.82\pm0.35$, which is consistent with the observed temporal slope at 1.1 sigma.

For the optical afterglow, we excluded data before 70 s from the fitting process as they were observed during the prompt emission phase and thus may be
dominated by the tail of the prompt emission. The best fit to the rest of the optical data required four segments. The first segment is an initial rise that peaks at 180 s. This is likely to be the onset of the afterglow and is discussed in more detail in Sec 4.4. In that section we focus on the decay segments. After the peak, a slope of $\alpha_{o,2}=0.91^{+0.03}_{-0.04}$ can be explained by the scenario of energy injection in a slow cooling interstellar medium (ISM) model with
$\upsilon_{m}<\upsilon_{o}<\upsilon_{c}$. We obtained a value of $q=0.73\pm0.04$, which is consistent with that derived from the X-ray energy injected decay
segment at 2$\sigma$ confidence level. Furthermore, the next segment with a temporal index of $\alpha_{o,3}=1.17^{+0.01}_{-0.01}$ is in agreement with the
$\alpha=3\beta/2$ at 3$\sigma$ confidence level, which also suggests that the afterglow ceases to be energy injected and enters the normal decay phase in which electrons are slow cooling in uniform medium. Other explanations are ruled out because the temporal indices derived using the spectral indices and the other closure relations are inconsistent with the
measured values at >3$\sigma$. The last optical decay segment breaks to a steeper decay at a time ($\sim$30000s) consistent, at 1$\sigma$, with the jet break
in the X-ray light curve. The optical decay slope for this segment is shallower than $-2.82\pm{0.35}$ derived using the X-ray spectral index, but is
consistent within 3$\sigma$ confidence level. As the jet break is a geometric effect, it should have an achromatic break time and the same post-break
decay index at all frequencies. For GRB 140629A, the break is achromatic in time, but the slopes of the post-break power law components are only marginally consistent; the decay index of the X-ray light curve is steeper than that in the optical. This has also been found for other GRBs such as GRB 050730 and GRB 051109A~\citep{2007MNRAS.380..374P}.

Overall, our analysis of the optical and X-ray light curves draws a consistent picture. The light curves are both produced by the blast wave jet impinging
on the constant density circumstellar medium in the slow cooling regime, where $\upsilon_{m}<\upsilon_{o}<\upsilon_{x}<\upsilon_{c}$. A long-lasting
central engine is still active after the prompt emission has vanished, which when it ceases, causes the light curve to enter the normal decay phase. At
$\sim$30000 s, an achromatic break is observed in the optical and X-ray light curves, which can be attributed to the jet break. The same process in both
bands supports the X-ray radiation and the optical radiation originating from a single component outflow.

\subsection{Physical model}
Following ~\citet{2015ApJ...806...15Z}, we fitted the multiwavelength data with a model based on numerical simulations to obtain further information about
the jet. This process is based on a 2D relativistic hydrodynamics (RHD) simulation, which assumes a jet with a top-hat Blandford-McKee profile~\citep{1976PhFl...19.1130B} that decelerates into a constant density medium. An ISM-type medium can be assumed as it has been found to explain the observations of most GRB afterglows~\citep{2001ApJ...560L..49P,2009ApJ...698...43R, 2011A&A...526A..23S} and is consistent with our analysis of GRB 140629A, as discussed in the previous section. Other assumptions of the model include that the radiation and dynamics of the collimated relativistic blast wave are assumed to be separate, and that the fraction of energy contained within the magnetic field at the front of the blast wave is low. The RHD simulation is performed with a relativistic adaptive mesh that employs a high-resolution adaptive mesh refinement (AMR) algorithm~\citep{2006ApJS..164..255Z}. This algorithm calculates the radiation transfer at a given observer time, angle, and distance along a line of sight~\citep{2012ApJ...749...44V,2013ApJ...767..141V}. The numerical model takes into account all the factors that can affect the shape of a jet break: (i) lateral expansion, (ii) edge effects, and (iii) off-axis effects. By fitting such a model to the optical and X-ray light curves, we are able to constrain some key physical parameters of the jet.

Because the data before the onset of the afterglow are still dominated by the prompt emission, we only model the data after 180\,s. We corrected
the optical data for extinction from the MW~\citep{2011ApJ...737..103S, 1998ApJ...500..525S} and the host galaxy, assuming the best fit host
extinction law from the SED fitting (SMC)~\citep[for details see the previous section,][]{1992ApJ...395..130P,2007MNRAS.377..273S,2010MNRAS.401.2773S}. 
We then converted the extinction corrected light curves to flux density at the central wavelength of the corresponding filter. For the X-ray light curve,
the galactic and host neutral hydrogen absorption was also corrected to get the intrinsic flux density at 1 keV.

The numerical modelling calculates the flux density at any frequency and observer time. The Markov Chain Monte Carlo method is used to determine the best
parameter values (i.e. the smallest $\chi^{2}$ value)~\citep{2016ApJ...833...88L,2017MNRAS.464.4624S}. The parameters determined include the total energy
$E_{tot,iso,53}$, the fraction of shock energy given to the electrons $\epsilon_e$, the fraction of shock energy given to the magnetic fields $\epsilon_B$,
the density of the medium $n$, the electron energy index $p$, the jet opening angle $\theta_{jet}$ and the observed angle $\theta_{obs}$. The starting
ranges for each parameter are $\theta_{jet}\in$ [0.045,0.5], $E_{tot,iso,53}\in$ [$10^{-10}$,$10^{3}$],
$n\in$ [$10^{-5}$,$10^{5}$], $p\in [2,4]$, $\epsilon_B\in$ [$10^{-10}$,1], $\epsilon_e\in$ [$10^{-10}$,1], and $\theta_{obs}$/$\theta_{jet}\in$ [0,1].
For more details, see~\citet{2015ApJ...806...15Z}.

\begin{table}
        \caption{Best fit parameters of the numerical simulation to the multiwavelength afterglow.}
        \label{Table4}
        \centering
        \begin{tabular}{c c c c}
                \hline
                \multicolumn{4}{c}{Modelling fitting}\\
                \hline
                Parameters & Value & Err (-) & Err (+)\\
                \hline
                $\theta_{jet}$ & 0.1171 & 0.0002 & 0.0061 \\
                $\log\,E_{tot,iso,53}$ &1.1414 &0.6226 &0.0675\\
                $\log\,n$ &4.6106 &0.5047 &  0.1875\\
                $p$ &2.0263 &0.0008 &0.0039 \\
                $\log\,\epsilon_B$ &-5.6730 &0.1616 &0.5201\\
                $\log\,\epsilon_e$ &-0.9974 &0.1347 &0.5294\\
                $\theta_{obs}$/$\theta_{jet}$& 0.5713 &0.0163& 0.0048\\
                \hline
        \end{tabular}
\end{table}

\begin{figure}
        \vspace{-1cm}
        \includegraphics[angle=0,scale=.55]{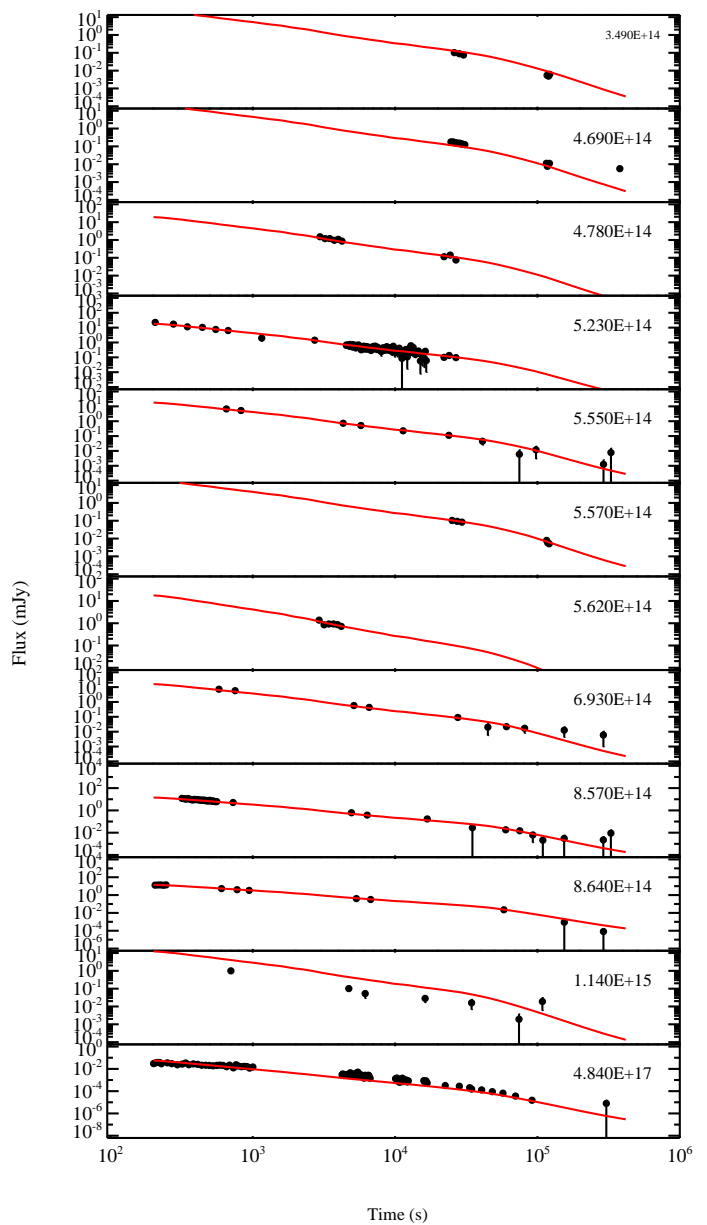}
        \caption{Best fit model determined from the numerical simulations overlaid on the observations at different wavelengths. The corresponding frequency is shown on the right corner in each panel in unit of Hz. The x-axis is the time since trigger in units of seconds. The observed flux density of each instrument is indicated on the y-axis in units of mJy. All data were corrected for MW and host galaxy absorption and extinction effects before modelling. Red solid lines represent the modelled light curves.}
        \label{multi-band lc}
        \vspace{-0.5cm}
\end{figure}

With these settings, the resulting best fit parameters and their uncertainties are listed in Table \ref{Table4}. The uncertainty on the parameters
are calculated at the $68\%$ confidence level in the local mode region. The best fit to each light curve for the different wavelengths is shown
in Figure \ref{multi-band lc} and the parameter distribution is given in Figure \ref{Modelling result}. In this case, the numerical model finds a
solution with best fit parameters of $\theta_{jet}\sim6.7^{\circ}$ and  $\theta_{obs}\sim3.8^{\circ}$, giving a total energy release of $1.4\times10^{54}$erg.
Since the modelling focusses on the effects of the jet break and based on a 2D RHD simulation, the energy injection is not taken into account, but the fit can still roughly describe most of the data. Furthermore the analytic approach ($p=2.8\pm0.3$) and the simulation
both have $p$ values that agree at 3$\sigma$. We find the opening angle to be typical of GRBs~\citep{2015ApJ...806...15Z,2009ApJ...698...43R}. The
relative off-axis angle, $\theta_{obs}/\theta_{jet} = 0.57$, is also consistent with the distribution, which peaks at 0.8, given in \cite{2015ApJ...806...15Z}.
We also obtained a high circumstellar density value that suggests this burst originated in a dense environment. In addition, our value for $\log\,\epsilon_B\sim$-5.7 is consistent with the modelling result from \citep{2018ApJ...860....8X}.

\subsection{Jet angle and empirical relation}

\begin{figure}
 \centering
  \includegraphics[width=17cm,angle=0,scale=0.6]{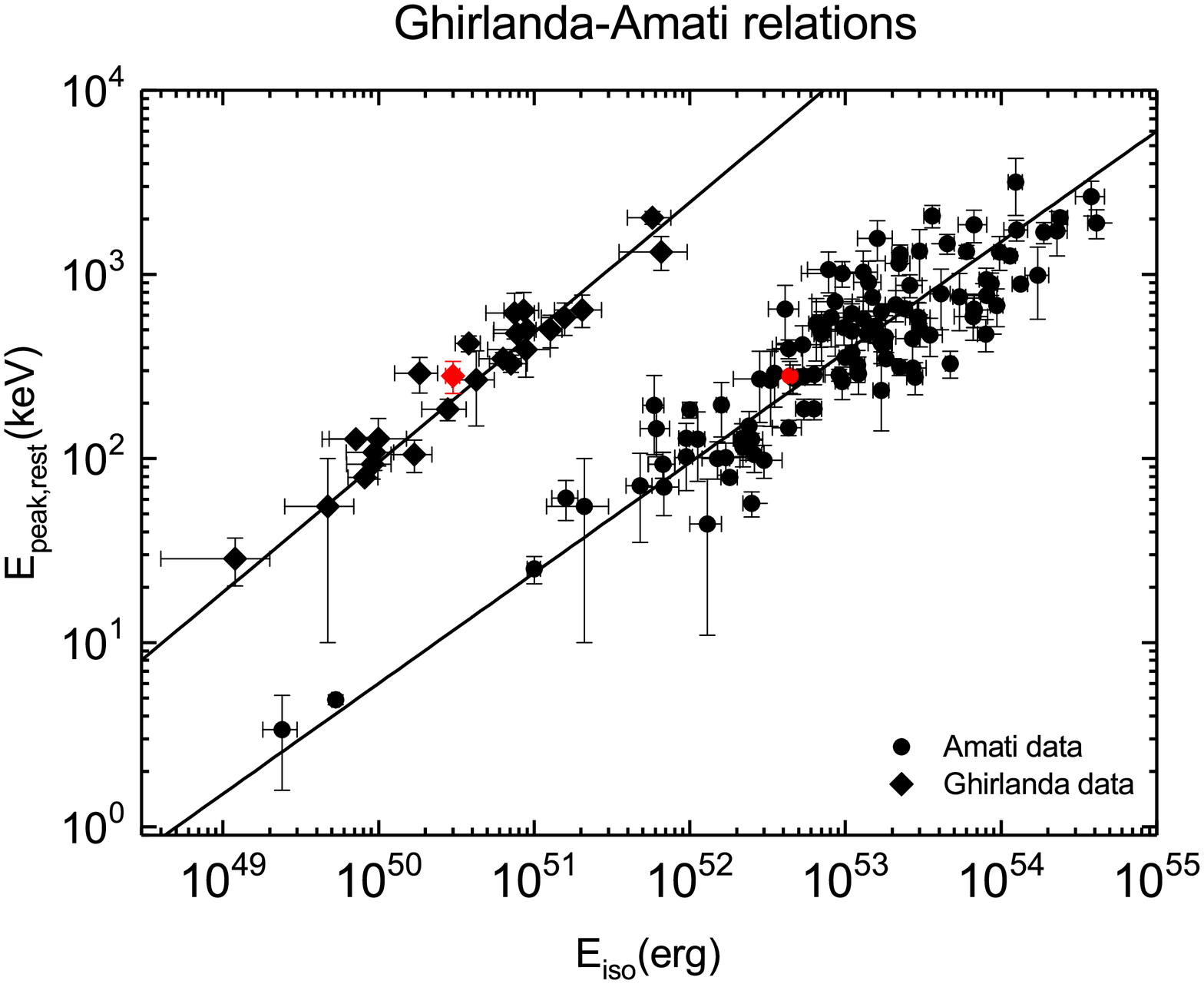}
 \caption{Location of GRB 140629A (red point) with the Ghirlanda (black prism) and Amati relation (black circle) derived from other typical GRBs~\citep[data from][]{2007A&A...466..127G,2008MNRAS.391..577A,2009A&A...508..173A}. The two straight lines
indicate the two empirical relations.}
 \label{relation}
 \vspace{-2.5ex}
\end{figure}

The $E_{r,iso}$ is the energy in $\gamma$-rays calculated assuming that the
emission is isotropic. The collimation corrected energy is calculated following
\begin{eqnarray}
 E_{\gamma} & = & E_{\gamma,\rm iso} f_{\rm b}=  E_{\gamma,\rm
iso}\left(1-\cos\theta_{\rm j} \right)
,\end{eqnarray}
where $f_{\rm b}$ is the collimation correction factor. 
For GRB 140629A, from the high-energy emission, we determined the isotropic rest-frame energy to be $E_{\gamma,\rm iso}=4.4\times10^{52}$\,erg and the observed $E_{peak}$= $86\pm17$\,keV. The peak of the energy spectrum in the rest frame is $E_{p,rest}$=$E_{peak}\times(1+z)$= $281\pm55$ keV. The $E_{\gamma,\rm iso}$ and $E_{p,rest}$ of this burst lie within the distribution of Amati correlation as shown in Figure \ref{relation}~\citep{2002A&A...390...81A,2008MNRAS.391..577A,2009A&A...508..173A,2012MNRAS.421.1256N}.

\begin{figure}
        \centering
        \includegraphics[width=17cm,angle=0,scale=.55]{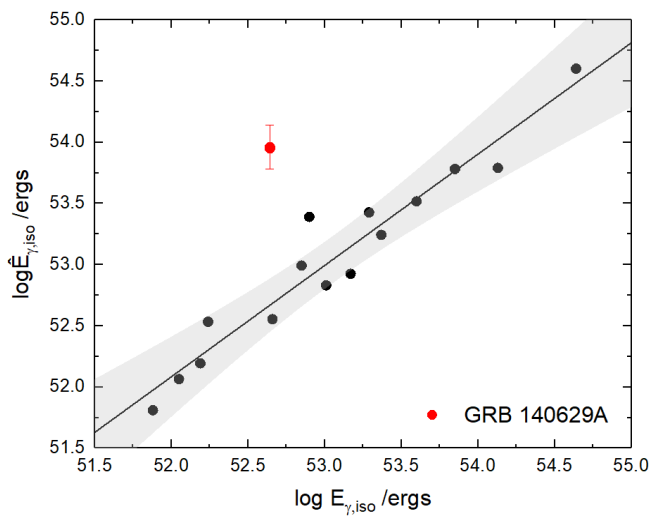}
        \caption{Location of GRB 140629A (red point) with the Liang-Zhang relation (black circle) derived from other typical GRBs~\citep[data from][]{2005ApJ...633..611L}. $\hat{\rm{E}}_{\gamma,iso}$ represents the calculated energy with the Liang-Zhang relation. The straight line indicates the empirical relations and the grey zone corresponds to the 3$\sigma$ confidence level.}
        \label{relation2}
\end{figure}

From the numerical modelling, we obtained a jet opening angle, $\theta_{jet}=6.7^{\circ}$. The collimation corrected energy is $E_{\gamma}=3.0 (\pm0.3)\times10^{50}$erg. Together with $E_{p,rest}$, this burst is also consistent with the Ghirlanda relation~\citep{2004ApJ...616..331G,2007A&A...466..127G} also shown in Figure \ref{relation}. GRB 140629A is denoted with a red point on both these empirical relations. The bootstrap method is used to estimate their errors. We also tested the relation between $E_{\gamma,\rm iso}$, $E_{p,rest}$ and $t_{b,rest}$ (the jet break time) known as Liang-Zhang relation~\citep{2005ApJ...633..611L}, but GRB 140629A is inconsistent with this correlation, shown in Figure \ref{relation2}. It is unclear why this GRB appears to be an outlier of the Liang-Zhang relation; it could be due to selection effects relating to the GRB prompt emission.

\subsection{Early optical rise}
Both MASTER and UVOT observed a peak at $\sim$180 s (see Figure \ref{all LC}). There are several explanations for this rise based on physical mechanisms
and geometric scenarios, which include the passage of the peak synchrotron frequency through the observing band, the reverse shock, decreasing extinction
with time, an off-axis jet, two component outflows, and the onset of the forward shock in the case of an isotropic outflow (\citet{2009MNRAS.395..490O}).
We discuss each of these options in turn.

The peak synchrotron frequency of the forward shock $\nu_{m,f}$ is expected to cross from blue to red frequencies, producing a chromatic peak which evolves at $t^{2/3}$. If the peak were due to the crossing of $\nu_{m,f}$ across the optical bands, the spectrum after peak is expected to be consistent with  $\nu^{-(p-1)/2}$ for $\nu_{m}<\nu<\nu_{c}$~\citep{1998ApJ...497L..17S}. The effective frequency of the MASTER clear filter and the UVOT white filter are
$5746 \AA$ and $3469 \AA$ respectively \citep{2008MNRAS.383..627P,2012ExA....33..173K}, thus the MASTER clear filter is the redder filter. Using the central
wavelengths of the white and clear filters converted to frequency and assuming  $p$=2, we predicted the peak in the bluer filter should appear 54 s earlier than the peak in the redder filter. We therefore selected the optical data from the MASTER clear filter and UVOT white filter between 90 s and 600 s and fit with a smooth broken power law model. The results are shown in Table \ref{Table5}. The measured difference between the peak times of the two filters is only 9 s and are the same within a 1$\sigma$ confidence level. This is inconsistent with the predicted peak time difference by 4$\sigma$. This therefore makes it unlikely that the passage of $\nu_{m,f}$ is the cause of the peak in the optical filters.

\begin{table}
        \caption{Fitting result from the first optical bump in two filters.}
        \label{Table5}
        \centering
        \begin{tabular}{c c c c }
                \hline
                Ins. & Slope1  & Peak time & Slope2\\
                \hline
                MASTER&  $-0.87_{-0.15}^{+0.16}$ &$170.0_{-11.8}^{+13.6}$&$1.09_{-0.13}^{+0.12}$\\
                UVOT&  $-1.01_{-0.10}^{+0.10}$ &$179.9_{-3.0}^{+3.4}$&$0.94_{-0.02}^{+0.03}$\\
                \hline
        \end{tabular}
\end{table}

For the reverse shock model, we just considered the constant density medium as this is consistent with the results of Section 4.1. In this case, when the peak synchrotron frequency of the reverse shock $\nu_{m,r}$ is lower than the optical wavelength, $\nu_{m,r}<\nu_{opt}$, the light curve is expected to decay after the peak with $\alpha=(3p+1)/4$~\citep{2003ApJ...595..950Z} with $p\sim2$ in this case, $\alpha\sim-1.75$. On the contrary, if $\nu_{m,r}>\nu_{opt}$, then the temporal index after the peak should be $\alpha\sim-0.5,$ which is followed by a decay of $\alpha=(3p+1)/4$. The slope after the early optical peak is inconsistent (>3$\sigma$ confidence) with both these scenarios for GRB 140629A. 

Another option to produce the rise could be dust destruction. An initially high level of dust could cause optical extinction, as this dust is destroyed by the radiation from the GRB, a chromatic peak is produced with different rise indices for the different filters \citep{2008A&A...483..847K}. As dust affects the bluest filters more strongly, the redder filters rise less steeply compared to the blue filters. While the redder MASTER filter has a shallower rise compared to the bluer UVOT filter, the slopes are consistent within 1$\sigma$ and we do not consider this to be a likely cause of the optical rise.

In the forward shock model, a peak is observed when the jet ploughs into the external medium. It is expected to produce an achromatic rise with $\alpha\sim1$ in the thick shell case with a constant density medium \citep{1999ApJ...520..641S,2002ApJ...570L..61G}. This is consistent with the rising slopes given in Table 6. We can exclude more complex jet geometry such as off-axis viewing and two component outflows. If the observer's viewing angle is larger than the
half-opening angle of the jet a rise is produced when the Lorentz factor $\varGamma$ decrease to $(\theta_{obs}-\theta)^{-1}$
\citep{2002ApJ...570L..61G,2005ApJ...630.1003G}. However, the modelling result shows that the observer angle is smaller than the half-opening angle, thus
this explanation can be excluded. Also the two-component outflow can be ruled out because we find that the afterglow can be explained by a single component
outflow in Sec 4.1. Thus, the achromatic peak and consistent slope make the forward shock the most likely option for GRB 140629A early optical bump.

\subsection{Initial bulk Lorentz factor}
The initial Lorentz factor ($\varGamma_{0}$) is an important parameter describing the initial parameters of the jet. A common way to estimate the initial
Lorentz factor is to use the peak time of the early afterglow light curve. The peak time determines the deceleration time of the external forward shock
and occurs when roughly half of the blast wave jet energy is transferred to the surrounding medium, as predicted in the blast wave jet
model~\citep{1999ApJ...519L..17S,2007ApJ...655..973K}. At this point, the Lorentz factor is half that of the $\varGamma_{0}$.
For a constant density medium, the initial Lorentz factor can be expressed as
\begin{equation}
   \varGamma_{\rm 0}=2.0 \left[\frac{3E_{\rm \gamma,iso}(1+z)^{3}}{32{\pi}nm_{\rm
p}c^{5}{\eta_{\gamma}}t_{\rm p}^{3}}\right]^{1/8},
   \label{eq:dec}
\end{equation}
where $z$ is the redshift, $n$ is the density of the external medium, $m_{p}$ is the proton
mass, $c$ is the speed of light, $\eta_{\gamma}$ is the radiation efficiency, and 
$t_{p}$ is the peak time of the afterglow onset bump~\citep{1999ApJ...519L..17S, 2010ApJ...725.2209L}. In the optical data, we found the early onset bump peaks at $\sim$180 s after trigger. Using the parameters, $n$ and $\eta_{\gamma}$, obtained from the modelling,  the initial Lorentz factor for GRB 140629A is $\varGamma_{0}=118\pm 5$. \citet{2012ApJ...751...49L} corrected the coefficient to 1.4 by integration of blastwave dynamics before the deceleration time. Using the revised equation, we obtained a Lorentz factor of $\varGamma_{0}=82\pm 4$, which is lower than the value of 315 obtained by \citep{2018ApJ...860....8X}. We likely got such different results because the parameters $\eta_{\gamma}$ and $n$ obtained from our modelling are one or two magnitudes higher than those used by \citet{2018ApJ...860....8X}. For GRB 140629A, the radiative efficiency, defined as $\eta_{\gamma}$=$E_{\rm \gamma,iso}$/$(E_{\rm \gamma,iso}$+$E_{\rm K,iso})$=$E_{\rm\gamma,iso}$/$E_{tot,iso}$, is 3.1\%; this is within the radiative efficiency distribution for long GRBs~\citep[see figure 10 in][]{2011ApJ...738..138R}.

\subsection{Properties of the optical polarisation}
The optical polarisation of GRBs provides additional clues to determine the structure and radiation mechanisms of the jet~\citep{2004ASPC..312..169C,2010arXiv1012.5101G}. Most GRB polarisation observations have been taken during the afterglow as the prompt emission is short lived. Both linear and circular polarisations have been found at optical wavelengths~\citep{1999A&A...348L...1C,2014Natur.509..201W}. 
In our observation of 140629A, we find an upper limit of $P<18\%$, which is consistent with the result from HOWPol \citep{2016MNRAS.455.3312G}.
Such a low degree of linear polarisation implies this burst is weakly polarised. This is considered to be confirmation that the dominant afterglow
emission mechanism is synchrotron radiation. Moreover, the polarisation measurement suggests an average dust-to-gas ratio in the GRB host galaxy
along the line of sight lower than our Galaxy \citep{2004AJ....128.1942K}, which is consistent with our findings in Section 4.7. It has been
proposed that the polarisation light curves have varied trends for various jet structures, especially at the jet break
time~\citep{2004MNRAS.354...86R,2010arXiv1012.5101G}. However, our polarisation observations were taken before the jet break and therefore we cannot use them to constrain the jet models for this burst.

\subsection{Properties of the host galaxy and environment}
The optical to X-ray SED at 9350\,s gives the $N_{H}$ along the line of sight as 7.2$\times 10^{21}$cm$^{-2}$, which is higher than that of our Galaxy
($N_{H}^{MW}$=9.3$\times$ $10^{19}$cm$^{-2}$ ) by two orders of magnitude. In addition, the intrinsic $E_{B-V}$ is 0.083 $\pm$ 0.009, which is also one
order of magnitude higher than that of our Galaxy. Therefore, the dust-gas ratio along the line of sight to GRB 140629A is $N_{H}/A_{V}=2.96\times10^{22}$cm$^{-2}$. This is lower than that of our Galaxy by one order of magnitude and is slightly lower than the mean value of $3.3\times10^{22}$cm$^{-2}$ for the
SMC extinction model from~\citealt{2010MNRAS.401.2773S}. This burst does not show any distinct feature in comparison with the other
typical GRBs in the $N_{H}-A_{V}$ plane~\citep[see Figure 9 in][]{2015MNRAS.449.2919L}.

\begin{figure}
        \centering
        \vspace{-5ex}
        \includegraphics[width=17cm,angle=0,scale=.6]{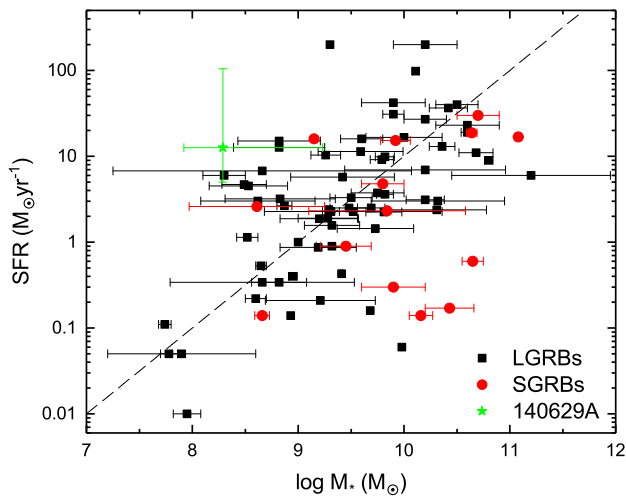}
        \caption{Plot of SFR vs. stellar mass for a sample of GRB hosts, inferred from template fitting to their photometric SEDs. The host of GRB 140629A is shown by a green star. Black squares and red dots represent the long burst and short burst hosts with SFRs measured from GHostS from 1997 to 2014~\citep{2006AIPC..836..540S,2009ApJ...691..182S}. The dashed line indicates a constant specific SFR of 1 Gyr$^{-1}$.}
        \vspace{-2.5ex}
        \label{SFR-M compare}

\end{figure}

\begin{figure}
        \centering
        \vspace{-2.5ex}
        \includegraphics[width=17cm,angle=0,scale=.6]{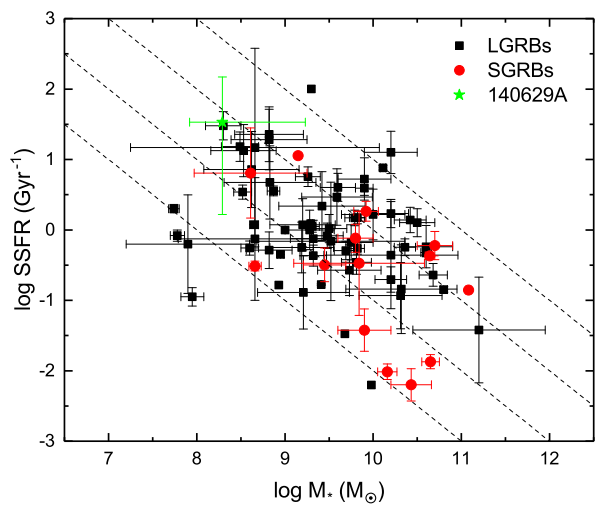}
        \caption{Plot of specific SFR vs. stellar mass for a sample of GRB hosts, inferred from template fitting to their photometric SEDs. The host of GRB 140629A is shown by a green star. Black dots indicate burst hosts with SFRs measured from GHostS from 1997 to 2014~\citep{2006AIPC..836..540S,2009ApJ...691..182S}. The dashed lines indicate the constant specific SFR of 0.1, 1, 10, and 100 Gyr$^{-1}$ from left to right.}
        \vspace{-2.5ex}
        \label{SSFR-M compare}
\end{figure}

The best fit of the host galaxy SED suggests the host galaxy has a SFR of log(SFR)=$1.1_{-0.4}^{+0.9}$ M$_\odot$ yr$^{-1}$.
Compared to other GRB host galaxies, the SFR is higher than the median value 2.5M$_\odot$ yr$^{-1}$~\citep{2009ApJ...691..182S}, but
within 2$\sigma$ of the distribution. The host galaxy is consistent with the SFR and stellar mass correlation for star-forming galaxies, known as
the star formation main sequence~\citep{2007ApJ...670..156D}, while it is at the edge of the distribution in the GRB sample shown in
Figure~\ref{SFR-M compare}~\citep{2009ApJ...691..182S}. This may indicate the mass of this galaxy is lower than other semi-SFR galaxies,
although the errors are fairly large. The specific SFR is higher than the average value of ~0.8Gyr$^{-1}$, but it follows the correlation
between the SSFRs and the stellar mass~\citep{2004A&A...425..913C,2009ApJ...691..182S}, as shown in Figure \ref{SSFR-M compare}.
The growth timescale in this case is lower than the Hubble time~\citep{2009ApJ...691..182S} at the burst distance, which suggests the galaxy
is in a bursty mode.

Damped Lyman-alpha ~\citep[DLA;][]{1986ApJS...61..249W,2005ARA&A..43..861W} systems trace the bulk of neutral hydrogen available for star formation
processes and are usually found in the lines of sight towards quasars ~\citep[QSOs;][]{2012A&A...540A..63N,2016MNRAS.456.4488S} and
GRBs~\citep{2009ApJS..185..526F,2013MNRAS.428.3590T}. Since GRBs are produced in star-forming regions, their sight-lines probe their surrounding
neutral environments within a few hundred parsecs of the sites of the bursts~\citep{2013A&A...549A..22V,2014A&A...564A..38D}. Hence, burst afterglow
absorption spectroscopy at z $\geq$1.8 (for which the Lyman-alpha absorption line is red-shifted out of the atmospheric blue cut-off) provides a
unique tool to investigate the amount of metals produced by vicinal star formation process. At a redshift of 2.276, GRB 140629A is therefore at
a suitable distance from which we were able to obtain the constituents of the GRB environment.

\begin{figure}
 \centering
  \includegraphics[width=17cm,angle=0,scale=.5]{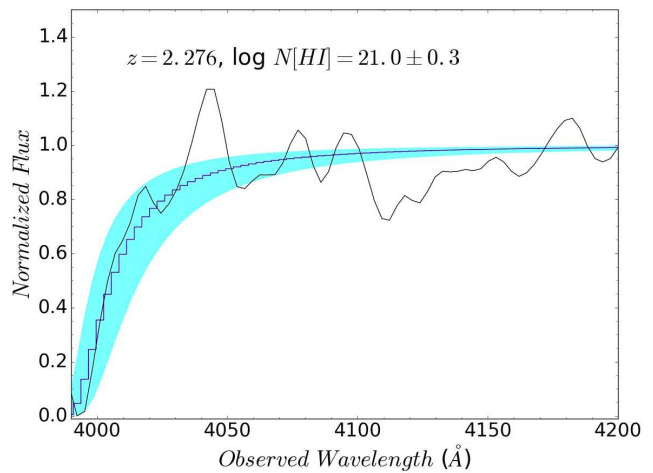}
 \caption{Voigt profile fit to the DLA in the spectrum of GRB 140629A. The figure shows
the data (black solid line) and the best fit damped profile (blue solid line) with its 1$\sigma$ confidence
interval (cyan area). }
 \vspace{-2.5ex}
 \label{DLA}
\end{figure}

\begin{figure*}
 \centering
 \includegraphics[width=17cm,angle=0,scale=0.75]{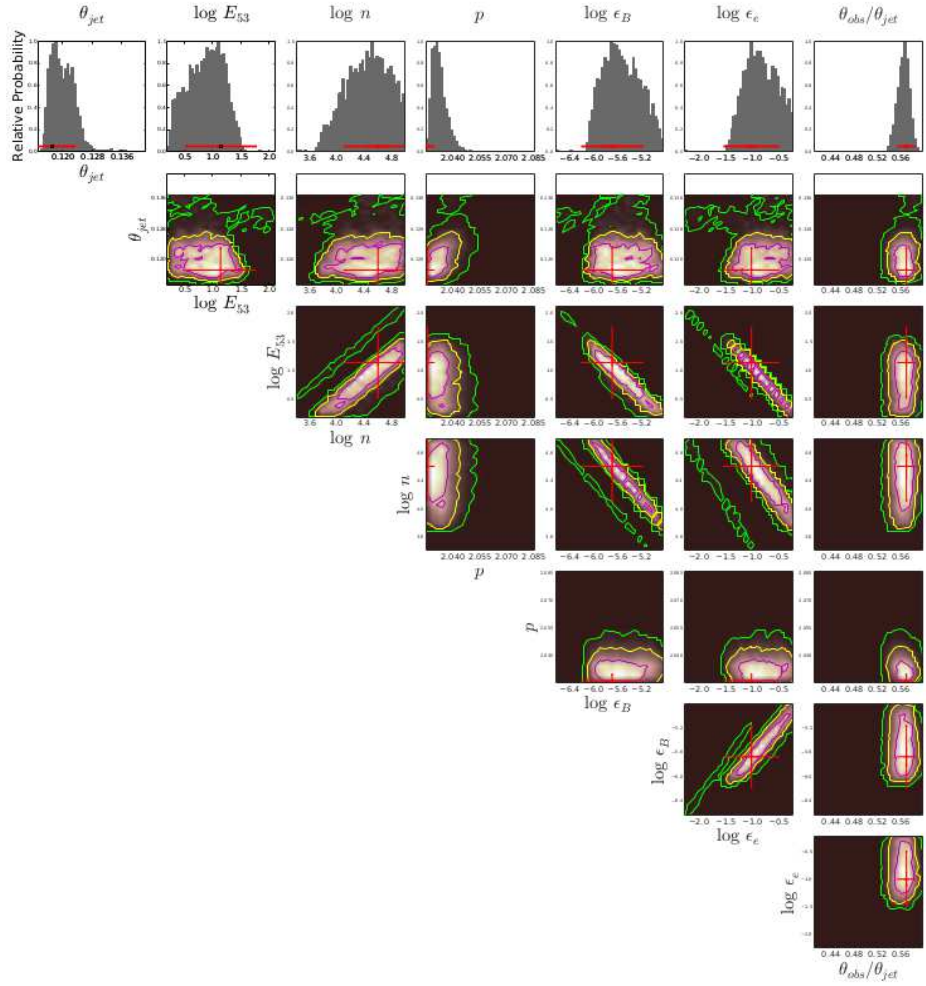}
 \caption{Triangle plot of the MCMC fitting to our simulation-based model. It shows the posterior distribution and the correlation between the parameters.}
 \label{Modelling result}
\end{figure*}

In order to investigate the neutral hydrogen content at the redshift of the host galaxy, we fitted the red damping wing of the Lyman-alpha absorption
with the Voigt profile using the same prescription and tools described in~\citet{2016MNRAS.456.4488S}, obtaining a column density value of
log $N_{H I}$=21.0$\pm$0.3, as shown in Figure \ref{DLA}. The large error in the fit mostly comes from the uncertainties in the continuum
determination due to the low S/N at the blue end of the spectra. In particular, the neutral hydrogen column density is the
characteristic indicator with which to distinguish if the host galaxy is a DLA system, by definition of
$N_{H I}\geq2\times10^{20}$cm$^{-2}$\citep{2005ARA&A..43..861W}. Therefore, the associated absorption system is technically classified as a DLA. 

The measured column density is lower than the peak value $N_{H I}=10^{21.5}$cm$^{-2}$ found in the GRB-DLAs distribution ~\citep{2009ApJS..185..526F},
but is still higher than the mean value of QSO-DLAs. Compared to other GRB-DLAs, this one does not show any properties distinct to the sample of
bursts in Figure 4 of ~\citet{2016ApJ...832..175T}. This value is unusual in the QSO-DLA sample, but frequently observed in GRB sight-lines, suggesting
once more that both samples are drawn from distinct populations.

In addition, we also identified strong high-ionisation lines, such as \ion{C}{IV}, \ion{Si}{IV,} and \ion{N}{V}, which are present at the redshift
of the absorber. In a previous analysis, it was found that the EWs of the GRB absorption feature are, on average, 2.5 times larger than
those in QSO-DLAs~\citep{2012A&A...548A..11D}. As shown in Table~\ref{TableEW}, those features in GRB 140629A are still consistent and even have an
excess on the high-ionisation lines, which is at least six times larger than the median value from QSO sample. Therefore, this burst also provides
evidence for EWs in GRB-DLA systems being larger than those in QSO-DLAs. That may imply that GRBs are produced inside the most luminous
regions of star-forming galaxies and that the light from the burst has to interact with much more host galaxy material. The \ion{N}{V} lines can be
used to trace collisionally ionised gas near long GRBs, since $N^{3+}$ has a high ionisation potential that makes the production of $N^{4+}$ difficult.
The cold \ion{N}{V} lines indicate that the GRB progenitor occurred within a dense environment $n\geq10^{3}$cm$^{-2}$~\citep{2008ApJ...685..344P}
within the photo-ionisation scenario. This indirectly supports the dense medium found through numerical modelling of GRB 140629A. Nevertheless, we
can neither constrain the distance of the \ion{N}{V} absorption to the progenitor, nor the metal abundance owing to the low signal-to-noise ratio
and resolution of the spectrum.

\section{Conclusions}
Thanks to the rapid response of several robotic telescopes and continued follow up by larger facilities, in this paper we are able, to present multiwavelength photometric
and spectroscopic observations of the long duration  GRB 140629A, providing a unique dataset on which to test models for this GRB. A detailed analysis of this burst was carried out to uncover the jet and host galaxy properties. This analysis is based on the data obtained by MASTER net, OSN, BOOTES, GTC, and BTA, as well as the public data from {\it Swift} and {\it Spitzer}. Optical spectroscopy obtained with BTA shows absorption features at a redshift of z=$2.276\pm0.001$ for this burst.

The signals in two orthogonal polarisations, measured by the MASTER telescope
of GRB 140629A, set an upper limit of 18\% at 1$\sigma$ confidence level
which implies that it is a weakly polarised burst and that synchrotron radiation dominates the afterglow emission.
Using the closure relations, we found that the afterglow in X-ray and optical bands can be well explained by a
long-lasting central engine which produces continued energy injection at the beginning. After the end of energy injection,
the normal decay phase is observed in both bands. The scenario in which a blast wave jet expands in a constant density ISM with slow cooling electrons, in the range $\upsilon_{m}<\upsilon_{o}<\upsilon_{x}<\upsilon_{c}$, can describe this burst
well during the phases with and without energy injection. We identify the final X-ray break at 31000 s  as a jet break. This break is achromatic and is observed in the optical at the same time and has break times consistent at 1$\sigma$. The afterglow is well explained by a single component outflow. 

We also attempted to model the broadband data with a blast wave jet model based on the prescription of~\citet{2015ApJ...806...15Z}.
The modelled result shows that this burst has a total energy release of $1.4\times10^{54}$ergs with an opening angle of
$6.7^{\circ}$ viewed $3.8^{\circ}$ off-axis. In addition, a high circumstellar density is obtained from modelling
and is also inferred indirectly from the identification of a high ionisation line (\ion{N}{V}). 

After correcting for redshift and the opening angle, for GRB 140629A we find the peak energy in the rest frame
and collimation-corrected energy are consistent with the Ghirlanda and Amati relations but not with the Liang-Zhang
relation. The optical light curve displays a peak, which we identified as the afterglow onset produced by the forward
shock which is the $\varGamma_{0}$ indicator. The onset is found at 181 s and indicates an initial Lorentz factor of 82-118.

Based on analysis of the host galaxy photometry, a low mass galaxy template with a SFR of
log(SFR)=$1.1_{-0.4}^{+0.9}$ M$_\odot$ yr$^{-1}$ at an age of  $1.14_{-0.35}^{+1.03}$Gyr is obtained. This result implies
the host galaxy is consistent with the star formation main sequence in a star-forming galaxy. Fitting the spectroscopy
at ~4000$\AA$ with a Voigt profile, a neutral hydrogen density log$N_{H I}=21.0\pm0.3$ derived indicates that we detect
a DLA system in the GRB host galaxy.

\begin{acknowledgements}
Acknowledge the support by the program of China Scholarships Council (CSC) under the Grant No.201406660015. We also acknowledge support from the Spanish MINEICO ministry and European FEDER funds AYA-2015-71718-R. SRO gratefully acknowledges the support of the Leverhulme Trust Early Career Fellowship. RS-R acknowledges support from ASI (Italian Space Agency) through the Contract n. 2015-046-R.0 and from European Union Horizon 2020 Programme under the AHEAD project (grant agreement n. 654215).
MASTER equipment is supported by Lomonosov MSU Development Program and by Moscow Union OPTIKA. VL,EG, NT, VK are supported by BRICS RFBR grant 17-52-80133.  NB (Tunka shared core facilities) are supported unique identifier RFMEFI59317X0005. B.-B.Z. acknowledges support from the National Key Research and Development Program of China (2018YFA0404204), and NSFC-11833003. S.B.P. acknowledges BRICS grant DST/IMRCD/BRICS/Pilotcall/ProFCheap/2017(G) for this work. I.D. acknowledges L. Piro his invitation and financial support to visit and work at IAPS (Rome). We also acknowledge the use of the public data from the $\it Swift$ data archive. We thank the excellent support form the GTC staff which is located at Observatorio del Roque de los Muchachos at Canary Islands (Spain). Thanks to the data support by NASA with $\it Spitzer$ Space Telescope.
Finally, we want to thank the anonymous referee for his/her comments, which have substantially improved the manuscript.
\end{acknowledgements}

%%%%%%%%%%%%%%%%%%%%%%%%%%%%%%%%%%%%%%%%%%%%%%%%%%%%%%%%%%%%%%%%%%%%%%%%%%%%%%%
%------------------------------------------------------------------------------
\bibliographystyle{aa} % style aa.bst
\bibliography{GRB140629Adraft}

-------------------------------
\newpage
\begin{center}
\topcaption{Photometric observations at the GRB 140629A field at optical
wavelengths. No correction for galactic extinction is applied.}
%\bottomcaption{This is the data}
\begin{supertabular}{|c|c|c|c|c|c|}
\hline%\noalign{\smallskip}
Ins.&Band & T-T$_0$ (s)   &Exp(s)  &  Mag  &Err\\
\hline
UVOT & $v$ & 643 & 10 & 15.30 & 0.09 \\
UVOT & $v$ & 817 & 10 & 15.57 & 0.10 \\
UVOT & $v$ & 4219 & 100 & 17.72 & 0.14 \\
UVOT & $v$ & 5654 & 100 & 18.10 & 0.14 \\
UVOT & $v$ & 10933 & 453 & 18.99 & 0.10 \\
UVOT & $v$ & 23772 & 93 & 19.75 & 0.40 \\
UVOT & $v$ & 40923 & 205 & 20.74 & 0.65 \\
UVOT & $v$ & 74215 & 454 & >22.2 &  nan \\
UVOT & $v$ & 91482 & 6229 & 22.19 & 1.54 \\
UVOT & $v$ & 142800 & 11769 & >22.1 &  nan \\
UVOT & $v$ & 265545 & 25697 & >23.9 &  nan \\
UVOT & $v$ & 328350 & 54 & >21.9 &  nan \\
UVOT & $b$ & 569 & 10 & 15.70 & 0.06 \\
UVOT & $b$ & 742 & 10 & 15.94 & 0.07 \\
UVOT & $b$ & 5039 & 100 & 18.45 & 0.08 \\
UVOT & $b$ & 6474 & 100 & 18.76 & 0.10 \\
UVOT & $b$ & 27297 & 281 & 20.45 & 0.20 \\
UVOT & $b$ & 44612 & 233 & 22.08 & 1.43 \\
UVOT & $b$ & 56976 & 3651 & 22.00 & 0.38 \\
UVOT & $b$ & 80901 & 364 & 22.26 & 0.85 \\
UVOT & $b$ & 142563 & 11702 & 22.59 & 1.21 \\
UVOT & $b$ & 264961 & 25578 & 23.41 & 2.01 \\
UVOT & $b$ & 328171 & 42 & >21.5 &  nan \\
UVOT & $u$ & 313 & 5 & 14.52 & 0.08 \\
UVOT & $u$ & 323 & 5 & 14.57 & 0.08 \\
UVOT & $u$ & 333 & 5 & 14.69 & 0.08 \\
UVOT & $u$ & 343 & 5 & 14.56 & 0.08 \\
UVOT & $u$ & 353 & 5 & 14.65 & 0.08 \\
UVOT & $u$ & 363 & 5 & 14.77 & 0.08 \\
UVOT & $u$ & 373 & 5 & 14.89 & 0.09 \\
UVOT & $u$ & 383 & 5 & 14.72 & 0.08 \\
UVOT & $u$ & 393 & 5 & 14.81 & 0.08 \\
UVOT & $u$ & 403 & 5 & 14.77 & 0.08 \\
UVOT & $u$ & 413 & 5 & 14.86 & 0.09 \\
UVOT & $u$ & 423 & 5 & 14.92 & 0.09 \\
UVOT & $u$ & 433 & 5 & 14.89 & 0.09 \\
UVOT & $u$ & 443 & 5 & 14.96 & 0.09 \\
UVOT & $u$ & 453 & 5 & 14.90 & 0.09 \\
UVOT & $u$ & 463 & 5 & 15.05 & 0.09 \\
UVOT & $u$ & 473 & 5 & 15.01 & 0.09 \\
UVOT & $u$ & 483 & 5 & 15.13 & 0.10 \\
UVOT & $u$ & 493 & 5 & 14.94 & 0.09 \\
UVOT & $u$ & 503 & 5 & 15.07 & 0.09 \\
UVOT & $u$ & 513 & 5 & 15.14 & 0.10 \\
UVOT & $u$ & 523 & 5 & 15.12 & 0.10 \\
UVOT & $u$ & 533 & 5 & 15.20 & 0.10 \\
UVOT & $u$ & 543 & 5 & 15.25 & 0.10 \\
UVOT & $u$ & 553 & 5 & 15.23 & 0.10 \\
UVOT & $u$ & 717 & 10 & 15.44 & 0.08 \\
UVOT & $u$ & 4834 & 100 & 17.74 & 0.08 \\
UVOT & $u$ & 6269 & 100 & 18.27 & 0.10 \\
UVOT & $u$ & 16707 & 116 & 19.13 & 0.15 \\
UVOT & $u$ & 34925 & 6 & >20.3 &  nan \\
UVOT & $u$ & 56063 & 3776 & 21.56 & 0.34 \\
UVOT & $u$ & 69374 & 5761 & 21.79 & 0.41 \\
UVOT & $u$ & 86665 & 6117 & 22.73 & 1.72 \\
UVOT & $u$ & 108800 & 305 & >23.1 &  nan \\
UVOT & $u$ & 142443 & 11665 & >22.7 &  nan \\
UVOT & $u$ & 264669 & 25515 & >23.1 &  nan \\
UVOT & $u$ & 328082 & 42 & >21.6 &  nan \\
UVOT & $white$ & 101 & 5 & 15.21 & 0.05 \\
UVOT & $white$ & 111 & 5 & 15.01 & 0.05 \\
UVOT & $white$ & 121 & 5 & 14.89 & 0.05 \\
UVOT & $white$ & 131 & 5 & 14.80 & 0.05 \\
UVOT & $white$ & 141 & 5 & 14.76 & 0.05 \\
UVOT & $white$ & 151 & 5 & 14.72 & 0.05 \\
UVOT & $white$ & 161 & 5 & 14.56 & 0.05 \\
UVOT & $white$ & 171 & 5 & 14.66 & 0.05 \\
UVOT & $white$ & 181 & 5 & 14.61 & 0.05 \\
UVOT & $white$ & 191 & 5 & 14.70 & 0.05 \\
UVOT & $white$ & 201 & 5 & 14.77 & 0.05 \\
UVOT & $white$ & 211 & 5 & 14.72 & 0.05 \\
UVOT & $white$ & 221 & 5 & 14.72 & 0.05 \\
UVOT & $white$ & 231 & 5 & 14.80 & 0.05 \\
UVOT & $white$ & 241 & 5 & 14.70 & 0.05 \\
UVOT & $white$ & 593 & 10 & 15.74 & 0.04 \\
UVOT & $white$ & 767 & 10 & 16.05 & 0.05 \\
UVOT & $white$ & 868 & 75 & 16.25 & 0.03 \\
UVOT & $white$ & 5244 & 100 & 18.54 & 0.05 \\
UVOT & $white$ & 6678 & 57 & 18.80 & 0.08 \\
UVOT & $white$ & 57887 & 225 & 21.69 & 0.29 \\
UVOT & $white$ & 142681 & 11739 & >24.4 &  nan \\
UVOT & $white$ & 265252 & 25641 & >27.0 &  nan \\
UVOT & $white$ & 328259 & 42 & >21.5 &  nan \\
UVOT & $uvw1$ & 692 & 10 & 17.52 & 0.33 \\
UVOT & $uvw1$ & 4629 & 100 & 20.01 & 0.41 \\
UVOT & $uvw1$ & 6064 & 100 & 20.71 & 0.68 \\
UVOT & $uvw1$ & 15801 & 450 & 21.39 & 0.60 \\
UVOT & $uvw1$ & 34019 & 450 & 22.02 & 0.96 \\
UVOT & $uvw1$ & 51655 & 5523 & >21.8 &  nan \\
UVOT & $uvw1$ & 68467 & 5757 & >23.6 &  nan \\
UVOT & $uvw1$ & 85759 & 5610 & >21.7 &  nan \\
UVOT & $uvw1$ & 107894 & 450 & 21.83 & 1.27 \\
UVOT & $uvm2$ & 677 & 10 & 20.05 &  0.86 \\
UVOT & $uvw2$ & 628 & 10 & 18.96 & 0.89 \\
OSN& V & 24852  & 600  & 19.83 & 0.11\\
OSN& V & 27020  & 600  & 19.96 & 0.11\\
OSN& V & 29186  & 600  & 20.07 & 0.11\\
OSN& V & 115713  & 600  & 22.63 & 0.19\\
OSN& V & 117851  & 600  & 22.97 & 0.22\\
OSN& V & 120677  & 600  & 23.11 & 0.30\\
OSN& I & 25781  & 600  & 18.99 & 0.20\\
OSN& I & 27944  & 600  & 19.14 & 0.20\\
OSN& I & 29931  & 600  & 19.33 & 0.20\\
OSN& I & 116628  & 600  & 22.17 & 0.35\\
OSN& I & 119433  & 600  & 22.29 & 0.33\\
OSN& I & 121592  & 600  & 22.03 & 0.46\\
OSN& R & 24537  & 300  & 18.89 & 0.15\\
OSN& R & 25469  & 300  & 18.89 & 0.15\\
OSN& R & 26393  & 300  & 19.03 & 0.15\\
OSN& R & 26706  & 300  & 19.04 & 0.15\\
OSN& R & 27632  & 300  & 19.06 & 0.15\\
OSN& R & 28557 & 300  & 19.09 & 0.15\\
OSN& R & 28871  & 300  & 19.17 & 0.15\\
OSN& R & 29798  & 300  & 19.22 & 0.15\\
OSN& R & 30723  & 300  & 19.28 & 0.15\\
OSN& R & 115404  & 300  & 21.89 & 0.27\\
OSN& R & 116320  & 300  & 21.97 & 0.26\\
OSN& R & 117235  & 300  & 22.32 & 0.34\\
OSN& R & 117543  & 2798  & 21.94 & 0.20\\
OSN& R & 120369  & 2130  & 21.90 & 0.22\\
OSN& R & 376511  & 5329  & 22.65 & 0.27\\
BOOTES&i&28937&1200&19.57&0.18\\
BOOTES&i&30413&1800&19.74&0.18\\
BOOTES&i&32305&1800&19.94&0.21\\
BOOTES&i&34310&1800&20.17&0.24\\
BOOTES&i&40041&2100&20.51&0.30\\
BOOTES&i&42288&2700&>19.97&nan\\
BOOTES&i&45521&2640&>19.65&nan\\
GTC&Sloan-g&$2.1\times10^{7}$&140$\times$3&>24.7&nan\\
GTC&Sloan-r&$2.1\times10^{7}$&90&>24.3&nan\\
GTC&Sloan-i&$2.1\times10^{7}$&90$\times$4&>24.6&nan\\
GTC&Sloan-g&$8.2\times10^{7}$&$150\times7$&25.01&0.20\\
GTC&Sloan-r&$8.2\times10^{7}$&$120\times7$&24.94&0.24\\
GTC&Sloan-i&$8.2\times10^{7}$&$90\times8$&24.71&0.32\\
Spitzer&3.6$\mu$m &$2.9\times10^{7}$ &$100\times72$&22.01&1.00\\
\hline%\noalign{\smallskip}
\multicolumn{6}{|c|}{MASTER-net}\\
\hline
Amur &P$\backslash$ &37&10 &14.26 &0.06\\
Amur &P$\backslash$ &72&10 &14.48 &0.06\\
Amur &P$\backslash$ &111&20 &14.06 &0.08\\
Amur &P$\backslash$ &151&30 &13.78 &0.13\\
Amur &P$\backslash$ &206&40 &13.86 &0.11\\
Amur &P$\backslash$ &277&50 &14.15 &0.07\\
Amur &P$\backslash$ &348&60 &14.61 &0.06\\
Amur &P$\backslash$ &443&80 &14.70 &0.07\\
Amur &P$\backslash$ &550&100 &15.05 &0.13\\
Amur &P$\backslash$ &672&120 &15.23 &0.17\\
Tunka &P$\backslash$ &1156&180 &16.50 &0.35\tablefootmark{a} \\
Tunka &C &2725&180 &16.85 &0.09        \\
Tunka &V &2924&180 &17.06 &0.08 \\
Tunka &R &2968&180 &16.62 &0.08 \\
Tunka &V &3172&180 &17.59 &0.13 \\
Tunka &R &3218&180 &16.88 &0.09 \\
Tunka &V &3426&180 &17.49 &0.12 \\
Tunka &R &3471&180 &16.87 &0.09 \\
Tunka &V &3691&180 &17.49 &0.12 \\
Tunka &R &3737&180 &17.11  &0.11 \\
Tunka &V &3941&180 &17.56 &0.13 \\
Tunka &R &3987&180 &16.98 &0.10 \\
Tunka &V &4188&180 &17.76 &0.16 \\
Tunka &R &4233&180 &17.23 &0.12 \\
Tunka &P- &4553&180 &17.87 &0.23 \\
Tunka &P$|$ &4553&180 &17.55 &0.14 \\
Tunka &P$|$ &4803&180 &17.72 &0.15 \\
Tunka &P- &4803&180 &17.79 &0.22 \\
Tunka &P- &5046&180 &17.81 &0.22 \\
Tunka &P$|$ &5047&180 &17.84 &0.17 \\
Tunka &P- &5287&180 &17.92 &0.24 \\
Tunka &P$|$ &5289&180 &17.98 &0.19 \\
Tunka &P$|$ &5531&180 &17.90 &0.17 \\
Tunka &P- &5533&180 &17.87 &0.23 \\
Tunka &P- &5778&180 &18.38 &0.32 \\
Tunka &P$|$ &5778&180 &18.29 &0.23 \\
Tunka &P$|$ &6023&180 &18.16 &0.21 \\
Tunka &P- &6025&180 &18.08 &0.27 \\
Tunka &P$|$ &6289&180 &18.19 &0.21 \\
Tunka &P- &6293&180 &18.13 &0.28 \\
Tunka &P- &6534&180 &18.25 &0.30 \\
Tunka &P$|$ &6534&180 &18.18 &0.21 \\
Tunka &P- &6775&180 &18.34 &0.31 \\
Tunka &P$|$ &6777&180 &18.41 &0.25 \\
Tunka &P$|$ &7016&180 &18.15 &0.21 \\
Tunka &P- &7019&180 &18.12 &0.27 \\
Tunka &P- &7257&180 &18.07 &0.27 \\
Tunka &P$|$ &7258&180 &18.13 &0.21 \\
Tunka &P- &7505&180 &18.25 &0.30 \\
Tunka &P$|$ &7507&180 &18.25 &0.22 \\
Tunka &P$|$ &7748&180 &18.26 &0.22 \\
Tunka &P- &7748&180 &18.45 &0.34 \\
Tunka &P- &7990&180 &18.99 &0.45 \\
Tunka &P$|$ &7991&180 &18.33 &0.24 \\
Tunka &P- &8233&180 &18.34 &0.31 \\
Tunka &P$|$ &8234&180 &18.34 &0.24 \\
Tunka &P$|$ &8490&180 &18.37 &0.24 \\
Tunka &P- &8491&180 &18.59 &0.36 \\
Tunka &P- &8733&180 &18.16 &0.28 \\
Tunka &P$|$ &8734&180 &18.23 &0.22 \\
Tunka &P- &8976&180 &18.30 &0.31 \\
Tunka &P$|$ &8977&180 &18.35 &0.24 \\
Tunka &P- &9228&180 &18.43 &0.33 \\
Tunka &P$|$ &9229&180 &18.40 &0.25 \\
Tunka &P$|$ &9471&180 &18.50 &0.26 \\
Tunka &P- &9471&180 &18.93 &0.43 \\
Tunka &P- &9711&180 &18.09 &0.27 \\
Tunka &P$|$ &9716&180 &18.59 &0.28 \\
Tunka &P$|$ &9952&180 &18.58 &0.28 \\
Tunka &P- &9953&180 &19.08 &0.47 \\
Tunka &P- &10189&180 &18.70 &0.38 \\
Tunka &P$|$ &10191&180 &18.78 &0.31 \\
Tunka &P- &10431&180 &19.07 &0.47 \\
Tunka &P$|$ &10432&180 &18.68 &0.30 \\
Tunka &P- &10682&180 &18.42 &0.33 \\
Tunka &P$|$ &10684&180 &18.75 &0.31 \\
Tunka &P- &10923&180 &19.21 &0.50 \\
Tunka &P$|$ &10924&180 &18.86 &0.33 \\
Tunka &P$|$ &11175&180 &18.79 &0.32 \\
Tunka &P- &11176&180 &20.13 &0.75 \\
Tunka &P- &11431&180 &18.87 &0.42 \\
Tunka &P$|$ &11431&180 &18.50 &0.26 \\
Tunka &P- &11686&180 &19.18 &0.49 \\
Tunka &P$|$ &11688&180 &19.09 &0.38 \\
Tunka &P- &11933&180 &18.62 &0.37 \\
Tunka &P$|$ &11934&180 &18.45 &0.26 \\
Tunka &P- &12173&180 &19.88 &0.67 \\
Tunka &P$|$ &12175&180 &18.57 &0.28 \\
Tunka &P- &12412&180 &18.61 &0.37 \\
Tunka &P$|$ &12413&180 &18.67 &0.29 \\
Tunka &P- &12652&180 &18.51 &0.35 \\
Tunka &P$|$ &12654&180 &18.82 &0.32 \\
Tunka &P$|$ &12905&180 &17.61 &0.14 \\
Tunka &P$|$ &13152&180 &18.32 &0.23 \\
Tunka &P- &13394&180 &18.29 &0.31 \\
Tunka &P$|$ &13396&180 &18.70 &0.30 \\
Tunka &P$|$ &13638&180 &18.87 &0.33 \\
Tunka &P$|$ &13882&180 &19.07 &0.37 \\
Tunka &P$|$ &14119&180 &18.78 &0.32 \\
Tunka &P$|$ &14347&180 &18.73 &0.31 \\
Tunka &P$|$ &14586&180 &18.50 &0.26 \\
Tunka &P$|$ &14819&180 &18.91 &0.34 \\
Tunka &P$|$ &15059&180 &20.17 &0.68 \\
Tunka &P$|$ &15301&180 &19.06 &0.37 \\
Tunka &P$|$ &15550&180 &19.81 &0.56 \\
Tunka &P$|$ &15799&180 &18.97 &0.35 \\
Tunka &P$|$ &16057&180 &19.91 &0.59 \\
Tunka &P$|$ &16302&180 &18.54 &0.27 \\
Tunka &P$|$ &16553&180 &20.11 &0.66 \\
Kislovodsk&C   &22078&1080\tablefootmark{b}&19.74 &0.13\\
Kislovodsk&R   &22078&1080\tablefootmark{c}&19.42 &0.22\\
Kislovodsk&C   &23985&1980\tablefootmark{d}&19.42 &0.13\\
Kislovodsk&R   &24382&1800\tablefootmark{e}&19.19 &0.22\\
Kislovodsk&C   &26759&1620\tablefootmark{f}&19.78 &0.13\\
Kislovodsk&R   &26759&2160\tablefootmark{g}&19.88 &0.22\\
\hline
\end{supertabular}
 \label{Table6}
\tablefoot{Optical data from different telescopes. 
	(Col. 1) Telescopes' name.
	(Col. 2) Filter used for observation.
	(Col. 3) The time interval between the middle of exposure and trigger time. 
	(Col. 4) Exposure time of observation.
	(Col. 5) Photometry data for GRB 140629A.
	(Col. 6) Error of the photometry data.		
For the UVOT observations, after 2000 s the exposure corresponds to the bin width rather than the exposure of individual images (see Section 3.1).  Photometry data for GRB 140629A by MASTER in the polarisers and R,V, C bands. The designation C~indicates white light that approximately corresponds to 0.2B+0.8R. The designation P$|$, P$\backslash$, P-~indicate polarisers orientated at 0$^\circ$, 45$^\circ$, 90$^\circ$ to the celestial equator, respectively. The absolute fluxes can be obtained using zero points from~\url{http://master.sai.msu.ru/calibration/}. All magnitudes are in Vega system except the GTC data.}
\tablefoottext{a}{\small Evening sky observation.}
\tablefoottext{b}{\small Coadd 6 frames.}
\tablefoottext{c}{\small Coadd 6 frames.}
\tablefoottext{d}{\small Coadd 11 frames.}
\tablefoottext{e}{\small Coadd 10 frames.}
\tablefoottext{f}{\small Coadd 9 frames.}
\tablefoottext{g}{\small Coadd 12 frames.}
\end{center}

\clearpage
\newpage

%%%%%%%%%%%%%%%%%%%%%

\end{document}